\documentclass[]{aastex631}

\usepackage{longtable}
\usepackage{amsmath}
\begin{document}

\title{The Diversity of Metal-Enrichment and Abundance Patterns at High Redshift: 
A Magellan Survey of Gas-rich Galaxies Traced by Damped Lyman-$\alpha$ Absorbers at $z\sim 5$}

\author{Jianghao Huyan}
\affiliation{Department of Physics \& Astronomy,
University of South Carolina,
Columbia, SC 29208, USA}

\author{Varsha P. Kulkarni}
\affiliation{Department of Physics \& Astronomy,
University of South Carolina,
Columbia, SC 29208, USA}

\author{Suraj Poudel}
\affiliation{Instituto de F\'isica, Pontificia Universidad Cat\'olica de Valpara\'iso, Casilla 4059, Valpara\'iso, Chile}
\affiliation{ Department of Physics and Astronomy, 
Texas Christian University,
 Fort Worth, TX, 76109, USA}

\author{Nicolas Tejos}
\affiliation{Instituto de F\'isica, Pontificia Universidad Cat\'olica de Valpara\'iso, Casilla 4059, Valpara\'iso, Chile}

\author{Celine P\'eroux}
\affiliation{European Southern Observatory, Karl-Schwarzschildstrasse 2, D-85748 Garching bei M{\"u}nchen, Germany}
\affiliation{Aix Marseille Universit\'e, CNRS, LAM (Laboratoire d'Astrophysique de Marseille) UMR 7326, 13388, Marseille, France}

\author{Sebastian Lopez}
\affiliation{Departamento de Astronomía, Universidad de Chile, Casilla 36-D, Santiago, Chile}

\begin{abstract}
{
A powerful technique to trace the signatures of the first stars is through the metal enrichment in concentrated reservoirs of hydrogen, such as the damped Lyman-$\alpha$ absorbers (DLAs) in the early universe. We conducted a survey aimed at discovering DLAs along sight lines to high-$z$ quasars in order to measure element abundances at $z$$>$4. Here we report our first results from this survey for 10 DLAs with redshifts of $\approx$4.2-5.0. We determine abundances of C, O, Si, S, and Fe, and thereby the metallicities and dust depletions. 
We find that DLA metallicities at $z$$>$4.5 show a wide diversity spanning $\sim$3 orders of magnitude. The metallicities of DLAs at 3.7$<$$z$$<$5.3 show a larger dispersion compared to that at lower redshifts. Combining our sample with the literature, we find a relatively smooth evolution of metallicity with redshift out to $z$$\sim$5.3, with a tentative ($\sim$2$\sigma$) indication of a slight rise in metallicity at 4.5$<$$z$$<$5.3. The relative abundances exhibit C enhancement for both metal-poor and metal-enriched DLAs. In addition, $\alpha$-element enhancement is evident in some DLAs, including a DLA at $z$=4.7 with a super-solar metallicity.  Comparing [C/O] and [Si/O] with model predictions, 4 DLAs in our survey seem consistent with a non-zero Pop III contribution (3 with $\gtrsim$30$\%$ Pop III contribution). Combining our sample and the literature, we  find the dust depletion strength and  dust-to-metal ratios to correlate positively with the total (gas+solid phase) metallicity, confirming the presence of metal-rich, dusty DLAs even at $\sim$1 billion years after the Big Bang.}

\end{abstract}


\section{Introduction} \label{sec:intro}
 The first stars are believed to have formed well within the first one billion years after the Big Bang from the primordial gas (possibly $<$200 million years after the Big Bang, i.e., at redshift $z\sim20$), and are referred to as Population III stars (Pop III stars). The launch of the James Webb Space Telescope (JWST) has brought a new perspective to the study of the formation of the first stars and the first galaxies.  Numerous high-redshift galaxies at $z > 5.5$ and even $z>10$ have been reported by the JWST Advanced Deep Extragalactic Survey (JADES) aimed at discovering high-redshift galaxies \citep{2023ApJ...952...74T,2024ApJ...962..124H,2023NatAs...7..622C,2023A&A...677A..88B,2024ApJ...970...31R,2024A&A...692A.184T}, putting constraints on stellar mass ($\sim 10^{7}-10^{9} M_{\odot}$) and star formation rates ($\sim 0.1-30$ $M_{\odot}$ {\rm yr}$^{-1}$) of high-redshift galaxies. While the highest redshift of the detected galaxies reaches $z \sim 14.3$, equivalent to $\sim 300$ million years after the Big Bang \citep{2024Natur.633..318C}, direct observations of the first stars and galaxies are still challenging. 

The nucleosynthetic signatures of the first stars can however be searched for by performing element abundance studies at high redshifts. Quasar absorption line spectroscopy enables robust measurements of metallicities and relative element abundances in the interstellar medium (ISM) and circumgalactic medium (CGM) of intervening galaxies situated along the lines of sight to the background quasars. For instance, the XQR-30 and its enlarged E-XQR-30 survey on the Very Large Telescope (VLT) provides a composite sample of high-quality spectra for over 30 quasars at redshifts $5.8<z<6.6$ \citep{2023MNRAS.523.1399D, 2024MNRAS.530.1829S} observed with X-Shooter. The XQ-100 survey determined the chemical abundances of DLAs at redshifts $z \sim 3-4$ observed with VLT/X-Shooter, and found [$\alpha$/Fe] ratios in multiple DLAs along a single quasar sightline show no difference relative to single DLAs matched in metallicity and redshift, indicating that the star formation history of DLA host galaxies is not affected by the presence of another nearby DLA \citep{2016MNRAS.463.3021B}. Low-ionization absorbers (LIAs), including O~{\sc i}, C~{\sc ii}, Si~{\sc ii}, Mg~{\sc ii} and Fe~{\sc ii}, trace the neutral cold gas in the CGM or IGM as their ionization potential is comparable to neutral hydrogen (13.6 eV). At high-redshift of z$\sim 5-6$, O~{\sc i} and C~{\sc ii} absorbers are investigated to study the chemical enrichment of the Universe during the Epoch of Reionization (EoR) \citep{2006ApJ...640...69B,2011ApJ...735...93B,2024MNRAS.530.1829S}. However, the LIAs without the detection of neutral hydrogen failed to measure the metallicities hence the metal evolution of cold gas in the CGM or IGM during the early cosmic time. To resolve this, cold gas with high column density of neutral hydrogen (e.g, damped Lyman-$\alpha$ absorbers, DLAs) is ideal for determination of metallicity in the gas, as its absorption lines of H~{\sc i} are prominent by showing damping wings.

The damped Lyman-$\alpha$ absorbers (DLAs) and sub-DLAs are reservoirs of cold absorbing gas containing amounts of neutral hydrogen and showing strong Lyman-$\alpha$ transitions in the spectra that indicate column densities of neutral hydrogen $N_{\rm H I} \ge 2\times 10^{20}$ and $10^{19} \le N_{\rm H I} < 2 \times 10^{20}$, respectively \citep{2003MNRAS.346.1103P}. The presence of the damping wings in the Lyman-$\alpha$ lines enables accurate determinations of H~{\sc i} column densities, thereby providing robust measurements of metallicity and element abundances at high redshifts. 

Unlike the low-redshift where stellar metal mass density is close to one half of total metal budget, at high redshift ($z \sim 4.5$) metal mass density of cool gas, such as DLAs, are estimated to contribute 94\% of total metal budget \citep{2024arXiv241119465D,2020ARA&A..58..363P}.
Measurements of element abundances in DLAs and sub-DLAs at $z \gtrsim 4.5$ are thus of great importance to investigate early chemical evolution of galaxies, determine the metal mass density at high-redshift end, and potentially to search for nucleosynthetic signatures of the first stars. The difficulty, however, lies in the dense Lyman-$\alpha$ forest at $z \gtrsim 4.5$, which makes it essential to use high-resolution spectroscopy to reliably measure the Lyman-$\alpha$ profiles and to identify and discern the metal lines. Obtaining adequately high-resolution, high-SNR spectra is especially challenging given the faintness of the high-$z$ quasars. 

The sample of high-resolution spectra of high-$z$ quasars has been growing slowly. 
Spectra obtained with the Echellette Spectrograph and Imager (ESI) on the Keck telescope have resolution of $\sim$33-44 km s$^{-1}$ FWHM, depending on the slit width used \citep{2012ApJ...755...89R,2014ApJ...782L..29R}.
Studies carried out using the X-Shooter spectrograph on the Very Large Telescope (VLT) have achieved resolutions of $\sim$55 km s$^{-1}$ \citep[XQ100;][]{2016MNRAS.463.3021B,2016A&A...594A..91L}, and $\sim$31 km s$^{-1}$ \citep[XQR30;][]{2023MNRAS.523.1399D}.
\cite{2018MNRAS.473.3559P,2020MNRAS.491.1008P} combine the spectra from ESI ($\sim$73 km s$^{-1}$), X-Shooter ($\sim$34-56 km s$^{-1}$), and the Magellan Echellette (MagE; $\sim$55 km s$^{-1}$) and Magellan Inamori Kyocera Echelle (MIKE; $\sim$13.6 km s$^{-1}$) spectrographs on Magellan telescope. Only 2 high-redshift quasars in these studies were observed at high spectral resolution [with MIKE by \citet{2020MNRAS.491.1008P}].

To improve this situation, in recent years, we have been performing high-resolution spectroscopy of quasars at $z \gtrsim 4.5$ using Magellan MIKE, which enables $\sim 2-3$ times higher resolution ($\sim$13.6 km s$^{-1}$) compared to X-Shooter, or $\sim 3-5$ times higher compared to ESI. Here we report a study of 10 (sub-)DLAs along the sight lines to 7 quasars in our sample. Section 2 describes the observations. Section 3 describes our measurement methodology. Section 4 discusses our results in the context of other DLA/sub-DLAs from the literature. Finally, section 5 summarizes our conclusions. 
Throughout this paper, we adopt the cosmological parameters $\Omega_{m}$=0.3, $\Omega_{\Lambda}$= 0.7, and $H_{0}$ = 70 km s$^{-1}$ Mpc$^{-1}$.

\section{Observations and Data Reduction} 

In order to study element abundances in DLAs and sub-DLAs, it is essential to measure both the Lyman-$\alpha$ line and the metal lines. In the absence of enough high-redshift DLAs/sub-DLAs confirmed with high-resolution spectroscopy, we targeted those high-$z$ quasars whose published low-resolution spectra \citep{2016ApJ...819...24W,2016ApJ...829...33Y,2019ApJ...871..199Y,2020MNRAS.491.1970W} suggested the presence of strong Lyman-$\alpha$ absorption features indicative of DLA/sub-DLA systems. We emphasize that the target selection was not dependent in any way on the strength of metal absorption lines. As described in the following section, our observations confirmed the presence of DLA/sub-DLAs in these sightlines.

\label{sec:obsv}
The observations were performed on the Magellan Clay telescope at the Las Campanas Observatory using the MIKE spectrograph during 2021-2023 as part of programs CN2021B-57 and CN2022A-61 (PI: Poudel), and NOIRLab programs 2023A-462536 and 2023B-190445 (PI: Huyan). Table 1 lists the targets and summarizes our observations. MIKE, as a double echelle spectrograph, provides a wavelength coverage of $\sim$4,900-10,000 \AA, and we focus on the red side covering $\sim$4,900-10,000 \AA.
The 1$\arcsec$ slit was used for all the observations, resulting in a resolving power of R$ \approx$22,000 on the red side. Since the targeted quasars are relatively faint ($i_{mag} \approx 18-20$), the observations were binned by a factor of 2 in both spatial and spectral axes in order to achieve a desirable signal to noise ratio (SNR, see \ref{tab:obs-tab}).
Exposures with an internal Th-Ar arc lamp were obtained for wavelength calibration; exposures with an internal quartz lamp were obtained for flat fielding. To get good flat-fields with the full CCD frame illuminated (including the regions between the Echelle orders) for robust pixel-to-pixel sensitivity corrections, ``milky flat" exposures with a diffuser slide inserted in the optical path were obtained.

The Python-based reduction pipeline CarPy \citep{2000ApJ...531..159K, 2003PASP..115..688K} was used for reducing the MIKE data including the science exposures, Th-Ar lamp exposures, quartz flats and milky flats. The pipeline includes the processes of bias subtraction, flat fielding, extraction of 1-dimensional spectra, wavelength calibration, and co-addition of multiple exposures. The extracted 1-dimensional spectra for the multiple orders were combined together to produce the final spectra. 

Continuum fitting was performed using a polynomial with the {\tt Linetools} spectral analysis code. The continuum-normalized spectra were examined to identify and measure absorption lines using {\tt VoigtFit} \citep{2018ascl.soft11016K}. 

\begin{table}[bt]
\caption{Target Observations.  
The SNR of the spectra are measured per resolution element at 1250\AA \ in the rest frame.
\label{tab:obs-tab}}
\begin{center}
\begin{tabular}{cccccclc}
\hline
\hline
QSO Name & RA(J2000) & Dec(J2000) & Magnitude  & $z_{QSO}$  & Exp. Time & Obs. Dates & SNR  \\
\hline

J0007-5701 & 00:07:36.56 & -57:01:51.8 & 17.21 (z) & 4.25 & 1200s$\times$3 & 20231107 & 47.2  \\
J0838-0440 & 08:38:32.31 & -04:40:17.47 & 19.62 (i) & 4.75 & 2700s$\times$4 & 20220303, 20220304 & 29.1  \\
J1129-0142 & 11:29:56.09 & -01:42:12.44 & 19.47 (i) & 4.87 & 2700s$\times$5 & 20220303, 20220304 & 39.0  \\
J1135-1526 & 11:35:30.4 & -15:26:10.25 & 19.19 (i) & 4.62 & 2700s$\times$4 & 20230418 & 36.4  \\
J1336–1830 & 13:36:14.74 & -18:30:43.29 & 18.78 (i) & 4.75 & 2400s$\times$4 & 20220303, 20220304 & 40.7  \\
J2202+1509 & 22:02:26.77 & +15:09:52.38 & 18.8 (i) & 5.07 & 1800s$\times$2 & 20210906 & 21.2  \\
J2328+0217 & 23:28:53.08 & +02:17:52.74 & 19.7 (i) & 4.86 & 2700s$\times$4 & 20210905, 20210906 & 27.6  \\

\hline
\hline
\end{tabular}
\end{center}
\end{table}

\section{Element Abundance Measurements} \label{sec:Abandances}

\subsection{Column Density Determinations}

Absorption lines in high-resolution spectra can be fitted with multi-component Voigt profiles to determine column densities of various elements, along with the Doppler parameters and velocities of each component. To this end, we performed the multi-component Voigt profile fitting of the detected absorption lines using the Python-based package {\tt VoigtFit} with Markov Chain Monte Carlo (MCMC) algorithm. 

\subsubsection{H~{\sc i} Lines}
\begin{figure}[ht!]
\centering
\includegraphics[width=1\linewidth, trim={50 90 50 125},clip]{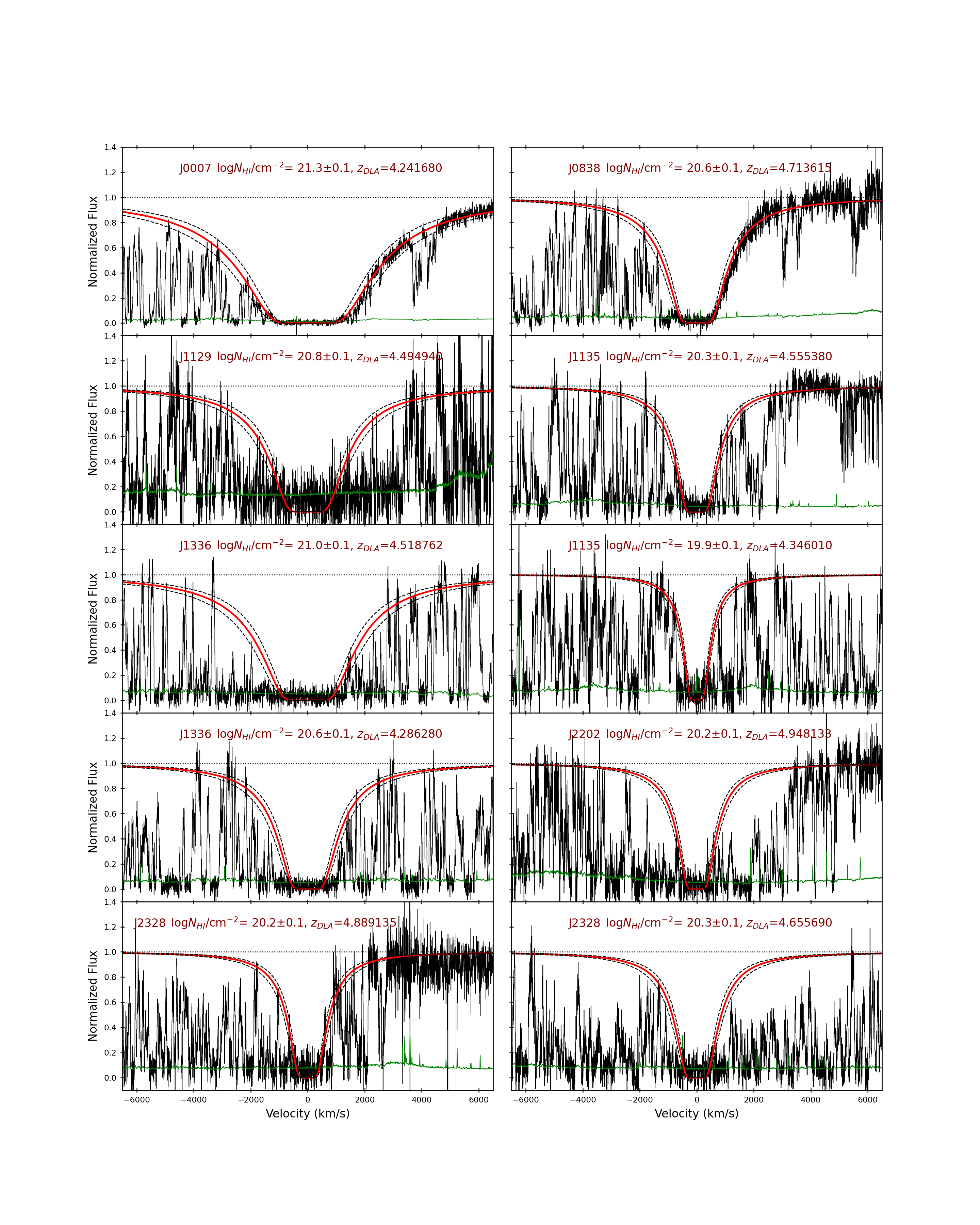}
\caption{Voigt profiles of Lyman-$\alpha$ lines for 10 DLAs in our survey. The fitted profiles are in solid red, normalized quasar spectra are in black, and the 1 $\sigma$ uncertainties in the normalized flux are in green. The dashed black curves show profiles corresponding to $\pm 0.1$ conservative uncertainties in ${\rm log}N_{\rm H I}$ values. The DLAs in the sightlines toward J0007-5701 and J0838-0440, the $z=4.555$ DLA toward J1135-1526, and the $z=4.889$ DLA toward J2328+0217 are located in the quasar proximity zones, with velocity separations $\Delta v \le 5000 \ {\rm km/s}$ relative to the quasar emission redshifts.
\label{fig:DLA-Lya}}
\end{figure}

The dense Lyman-$\alpha$ forest lines at high-redshift make the measurements of DLA profiles difficult because of the challenges in the determination of the quasar continuum. However, in most cases, the presence of damping wings is still clear, allowing reasonably robust determinations of the H~{\sc i} column densities. In cases where the Lyman-$\alpha$ forest blending is particularly severe, data in the continuum, the core of the line, as well as the highest points within the wings were used to constrain the H~{\sc i} column densities. We fitted Voigt profiles to the uncontaminated regions of Lyman-$\alpha$ absorption lines with log $N_{\rm H I}/(\mathrm{cm^{-2}})$ in the range of $19.3-21.8$, with increments of 0.05 dex per step. The best-fitting H~{\sc i} column densities are determined by lowest $\chi^2$ as defined by {\tt VoigtFit} and the conservative $\pm 1 \sigma$ uncertainties of 0.1 dex (which cover the uncertainties in continuum placement) are listed in Table 2. Fig. \ref{fig:DLA-Lya} shows the Voigt profile fits for the Lyman-$\alpha$ lines for both the best-fitting H~{\sc i} column densities and the $\pm 1 \sigma$ uncertainties.
\subsubsection{Metal Lines}

For the DLAs in our survey, we first selected metal lines that are not saturated and not blended with telluric lines or lines from absorption systems other than the DLA system of interest, and thereafter performed Voigt profile fitting on them in order to decide the velocity structure of the DLA absorbing system, and the Doppler parameters and column densities of the relevant metal ions in the confirmed components. The velocity structure and Doppler parameters thus determined were adopted to fit the other metal lines associated with the DLA, to determine the respective column densities. In general, our survey covers the absorption lines of C~{\sc ii}, O~{\sc i}, Si~{\sc ii}, S~{\sc ii}, and Fe~{\sc ii}. Table.\ref{tab:absorption-catlog} lists the fitting results for the DLAs from our survey, and Fig. \ref{fig:J0007_metals} to Fig. \ref{fig:J2328_metals-lowz} show the Voigt profiles for these metal lines.

\begin{table}[bt]
\caption{Observed Metal Abundances \label{tab:abs-sum}}
\begin{center}
\begin{tabular}{cccccccc}
\hline
\hline
QSO Name & $z_{DLA}$ & log$N_{HI}$/cm$^{-2}$ & [C/H]  & [O/H] & [Si/H] & [S/H] & [Fe/H] \\
\hline
J0007-5701 & 4.24168 & $21.30_{-0.10}^{+0.10}$ & $\ge -1.03$	& $\ge -0.84$	& $\ge -1.53$	& $-1.59_{-0.10}^{+0.10}$	& $-1.63_{-0.14}^{+0.15}$	\\
J0838-0440 &  4.713615 & $20.60_{-0.10}^{+0.10}$ & $\ge -1.75$	& $\ge -1.34$	& $\ge -0.91$	& $-0.44_{-0.10}^{+0.11}$	& $-0.67_{-0.35}^{+0.91}$	\\
J1129-0142 & 4.494940 & $20.80_{-0.10}^{+0.10}$ & $0.98_{-0.14}^{+0.18}$	& $-1.02_{-0.25}^{+0.22}$	& $-1.54_{-0.12}^{+0.14}$	& -	& -	\\
J1135-1526 & 4.555380 & $20.30_{-0.10}^{+0.10}$ & $-1.61_{-0.12}^{+0.14}$	& $-1.34_{-0.11}^{+0.12}$	& $-1.36_{-0.10}^{+0.10}$	& $-1.04_{-0.11}^{+0.11}$	& $-1.83_{-0.11}^{+0.11}$	\\
J1135-1526 & 4.346010 & $19.90_{-0.10}^{+0.10}$ & $-2.35_{-0.16}^{+0.28}$	& $-0.61_{-0.84}^{+0.51}$	& $-2.62_{-0.24}^{+0.18}$	& -	& -	\\
J1336-1830 & 4.518762 & $21.00_{-0.10}^{+0.10}$ & $-0.90_{-0.23}^{+0.19}$	& $-0.82_{-0.22}^{+0.19}$	& $-2.17_{-0.13}^{+0.14}$	& -	& $-2.78_{-0.11}^{+0.11}$	\\
J1336-1830 & 4.286280 & $20.60_{-0.10}^{+0.10}$ & $-1.21_{-0.43}^{+0.42}$	& $\ge -1.51$	& $-1.78_{-0.11}^{+0.13}$	& -	& $-2.33_{-0.11}^{+0.11}$	\\
J2202+1509 & 4.948650  & $20.20_{-0.10}^{+0.10}$ & $-2.00_{-0.14}^{+0.16}$	& $-2.17_{-0.12}^{+0.12}$	& $-1.95_{-0.11}^{+0.11}$	& -	& -	\\
J2328+0217 & 4.889080 & $20.20_{-0.10}^{+0.10}$ & $-2.92_{-0.17}^{+0.19}$	& $-1.51_{-0.44}^{+0.38}$	& $-2.59_{-0.25}^{+0.27}$	& -	& -	\\
J2328+0217 & 4.655690 & $20.30_{-0.10}^{+0.10}$ & $-1.46_{-0.59}^{+1.12}$	& $-0.84_{-0.77}^{+0.89}$	& $-1.69_{-0.11}^{+0.11}$	& -	& -	\\
J0025-0145 &  4.7372 & $20.48_{-0.15}^{+0.15}$ & $0.88_{-0.17}^{+0.17}$ & $0.71_{-0.16}^{+0.16}$ & $-0.56_{-0.35}^{+0.40}$	& -	& $-1.53_{-0.15}^{+0.15}$  \\
\hline
\hline
\end{tabular}
\end{center}
\end{table}

\begin{table}[bt]
\caption{Dust Depletions and Dust-to-Metal Ratios for 4 DLAs with both Si and Fe detections. [Zn/Fe]$_{exp}$ denotes the expected [Zn/Fe] ratio from [Si/Fe] ratio. $\delta_{X}$ (X=Si and Fe) denotes the dust depletion for element X. [Fe/H]$_{\rm tot}$ denotes [Fe/H] after dust depletion correction. $DTM_{Fe}$ denotes the dust-to-metal ratio based on Fe. $F_{*}$ is a factor representing the overall dust depletion strength introduced by \cite{2009ApJ...700.1299J}, and [M/H]${\rm intrinsic}$ is the intrinsic metallicity corrected for dust depletion. The calculations [following \cite{2018A&A...611A..76D} and \cite{2009ApJ...700.1299J}] are described in Sec. \ref{sec:dust-depletion} and \ref{sec:DTM}. \label{tab:DTM}}
\begin{center}
\begin{tabular}{cccccccccc}
\hline
\hline
QSO Name & $z_{DLA}$ & [Si/Fe] & [Zn/Fe]$_{exp}$  & $\delta_{Fe}$ & $\delta_{Si}$ & [Fe/H]$_{\rm 
 tot}$ & DTM$_{Fe}$ & $F_{*}$ & [M/H]$_{\rm instrinsic}$ \\
\hline

J0025+0145 & 4.7372 & $1.34_{-0.84}^{+0.41}$ & $2.20_{-1.74}^{+0.88}$ & $-2.79_{-2.19}^{+1.11}$ & $-1.42_{-1.10}^{+0.57}$ & $1.16_{-2.19}^{+1.12}$ & $0.9984$ & $1.35 \pm 0.27$ & $1.10 \pm 0.22$  \\
J1135-1526 & 4.5554 & $0.47_{-0.07}^{+0.07}$ & $0.43_{-0.16}^{+0.16}$ & $-0.55_{-0.21}^{+0.21}$ & $-0.30_{-0.11}^{+0.11}$ & $-1.28_{-0.24}^{+0.24}$ & $0.7182$ & $-0.32 \pm 0.19$ & $-1.45 \pm 0.18$  \\
J1336–1830 & 4.5188 & $0.61_{-0.11}^{+0.10}$ & $0.71_{-0.25}^{+0.23}$ & $-0.91_{-0.32}^{+0.29}$ & $-0.48_{-0.17}^{+0.15}$ & $-1.87_{-0.33}^{+0.31}$ & $0.8770$ & $1.10 \pm 0.22$ & $-0.61 \pm 0.23$  \\
J1336–1830 & 4.2863 & $0.59_{-0.11}^{+0.09}$ & $0.67_{-0.25}^{+0.21}$ & $-0.86_{-0.31}^{+0.27}$ & $-0.45_{-0.16}^{+0.14}$ & $-1.46_{-0.33}^{+0.29}$ & $0.8615$ & $0.46 \pm 0.39$ & $-0.96 \pm 0.45$  \\

\hline
\hline
\end{tabular}
\end{center}
\end{table}

\subsection{Ionization Corrections}
Atoms in the CGM can be ionized into different ionization states by high-energy photons from the background ultraviolet (UV) radiation, and from local sources within the absorbing galaxy. On the other hand, observations often allow measurements of column densities in only the dominant state of ionization. In such cases, measurements of element abundances require corrections accounting for the ions not accessed by the observations.
For neutral gas with high H~{\sc i} column densities such as DLAs, the ionization corrections are generally expected to be not significant as the neutral gas is self-shielded and most elements such as C, Si, S, and Fe are mostly in the singly ionized state. However, in absorbers with lower H~{\sc i} column densities, such as sub-DLAs and Lyman-limit systems (LLS), ionization corrections can be more significant. In our sample, the only sub-DLA where ionization corrections could be determined is the sub-DLA at $z_{abs}=4.889080$ in the sight line to J2328+0217, which shows absorption lines from multiple ionization states. 

We use the {\tt Cloudy} c23.01 plasma simulation code \citep{2023RMxAA..59..327C} to perform photoionization modeling in order to constrain the ionization parameter (defined as the ratio of the photon number density to the number density of neutral hydrogen atoms $U=n_{\gamma}/n_H$) of the absorbing gas, and thereby to estimate ionization corrections to the metallicity determinations. For the {\tt Cloudy} model of the sub-DLA along the slight line to J2328+0217, we treat the system as a single-structure slab and simulate its thermal, ionization, and chemical structure in the presence of the cosmic microwave background (CMB), cosmic rays, local interstellar radiation, and the background UV radiation at the redshift of the absorber. For the background UV radiation, we use the model for extragalactic radiation from \cite{2019MNRAS.484.4174K}, which adopted measurements of quasar emissivity, star formation rate density and dust attenuation in galaxies, and the distribution of H~{\sc i} gas in the IGM. 

We perform a grid of {\tt Cloudy} models with $-2.5<$ log $n_{HI}$ $<-2.3$ in steps of 0.01 dex. In these models, the ionization parameter is constrained to be $-1.09<$ log $U<-1.06$ by the observed C~{\sc ii} and C~{\sc iv} column densities 
(see Fig.\ref{fig:photonionization}). The photoionization correction to metallicity is estimated to be $<0.06$ dex based on element O, which is within $1\sigma$ error of the observational measurement and can be considered negligible. Table \ref{tab:abs-sum} lists the gas-phase element abundances derived from our observations.

\subsection{Dust Depletion Corrections}\label{sec:dust-depletion}
Determination of metallicities from the measured column densities is based on metals in the gas phase, but a part of the metals can be incorporated into the solid phase, i.e. condense into dust grains. These dust depletion effects are more significant for refractory elements that have higher condensation temperatures. A correction is necessary to account for these missing metals in order to determine the total (gas-phase + solid-phase) metallicities. \cite{2009ApJ...700.1299J} developed a method to estimate the dust depletion corrections by comparing the measured element abundances for different elements spanning a range of condensation temperatures (i.e., both refractory and volatile elements). This method was based on the measurements of 17 elements in 243 interstellar sight lines in the Milky Way to characterize the correlation between relative abundances and and dust depletion factor in terms of three parameters as 
$$ [X/H]_{\rm obs} - B_{X} + A_{X} \cdot z_{X} = [X/H]_{\rm intrinsic} + A_{X} F_{*} $$
where A$_X$, B$_X$ and z$_X$ are specific to and solved for each element from the sample, F$_*$ estimates the overall strength of dust depletion, $[X/H]_{\rm obs}$ is observed abundance of element X compared to solar and $[X/H]_{\rm intrinsic}$ is the intrinsic (dust-corrected) metallicity compared to solar after the correction for dust depletion. We applied this method to measurements for the absorbers in our survey for various elements to estimate the depletion factor and the intrinsic (dust-corrected) metallicities. Table 3 lists 
the dust depletion estimates.

\subsection{Estimation of Dust-to-Metal ratios}\label{sec:DTM}

We now investigate the dust-to-metal (DTM) ratios of the DLAs. To this end, we calculate the dust depletions based on element ratios relative to Fe, following the prescription of \cite{2018A&A...611A..76D}, and then derive the DTM of Fe.
The depletion of element X, indicating the logarithmic amount (in dex) of X observed (in the gas phase) relative to its total abundance is $\delta_{X} = [X/H]_{\rm obs} - [X/H]_{\rm tot} - \alpha_{X}$ where [X/H]$_{\rm obs}$ is the observed abundance of element $X$, $[X/H]_{\rm tot}$ is the total metallicity based on element $X$, and $\alpha_{X}$ is the intrinsic (nucleosynthetic) over- or under-abundance of X with respect to Fe. 

When a Zn measurement is available, the dust depletion can be estimated from the [Zn/Fe] relative abundance (which serves as a reliable tracer of overall dust depletion effect) as $\delta_{X} = A2_{X} + B2_{X} \times [Zn/Fe]$, where $A2_{X}$ and $B2_{X}$ are parameters estimated from the empirical fits to the depletion sequences \citep{2016A&A...596A..97D,2018A&A...611A..76D}. For example, for Fe, $A2_{Fe} = -0.01 \pm 0.03$ and $B2_{Fe} = -1.26 \pm 0.04$ \citep{2016A&A...596A..97D,2018A&A...611A..76D}. 

When Zn measurements are not available, the expected [Zn/Fe] can be estimated from the relative abundance ratio [X/Fe] of other elements, since $[Zn/Fe]_{exp} = ([X/Fe] - A1_{X})/(B1_{X} + 1)$, where $A1_{X}$ and $B1_{X}$ are 
parameters estimated from empirical fits between [X/Zn] and [Zn/Fe] \citep{2016A&A...596A..97D,2018A&A...611A..76D}. For example, for Si, $A1_{Si} = 0.26 \pm 0.03$ and $B1_{Si} = -0.51 \pm 0.06 $ for element Si. Since Zn measurements are not available for our DLAs, we use the observed [Si/Fe] to estimate [Zn/Fe]$_{exp}$, and thereby estimate the Fe depletion $\delta_{Fe} = A2_{Fe} + B2_{Fe} \times [Zn/Fe]$ using the above-mentioned values of $A2_{Fe}$ and $B2_{Fe}$. The DTM (defined as the fraction of the metal in the dust phase) for Fe can then be estimated from the depletion $\delta_{Fe}$ as $\mathrm{DTM}_{Fe} = 1 - 10^{\delta_{Fe}}$. Table 3 lists the Fe and Si depletions, dust depletion-corrected Fe-based metallicity, and the DTM$_{Fe}$ for the DLAs in our sample with Fe measurements.

\section{Discussion}
\subsection{Evolution of metallicity}
The evolution of metallicity reflects how galaxies evolve over cosmic time, which can be investigated by the global metallicity-redshift relationship. It was suggested that the mean metallicity decreases $\sim 0.20-0.30$ dex per unit redshift with the redshift range $0.5 < z < 5$ \citep{2003ApJ...595L...9P}. The low-redshift end $0.09<z<0.52$ of the metallicity-redshift relation was determined to have a slope of $-0.23\pm0.06$ per unit redshift based on the nearly undepleted element Zn \citep{2005ApJ...618...68K}. Subsequent studies \citep{2006MNRAS.370...43M,2012ApJ...755...89R,2016ApJ...830..158M} suggested a slope of $\sim$0.2 dex per unit redshift over the range $z < 4$. A sudden metallicity evolution at $z>4.7$ was proposed based on Si and Fe measurements for 16 DLAs, suggesting rapid metallicity enrichment after the reionization epoch \citep{2014ApJ...782L..29R}. However, other works based on measurements of the nearly undepleted elements O and S did not display a sudden decline at $z\sim4.7$ compared to predictions \citep{2018MNRAS.473.3559P,2020MNRAS.491.1008P}. This difference may stem from the use of different types of metals for determination of metallicity in these works, since refractory elements (such as Fe and Si) can be incorporated into dust grains and thus depleted from the gas phase, while volatile elements such as O and S show much weaker depletion. 

We calculate the metallicities for our sample of absorbers based on the nearly undepleted elements O and S, following the standard definition 
$$[X/H] = {\rm log}(N_{X}/N_H) - {\rm log}(X/H)_{\odot}$$
where $N_{X}$ represents the column density determined by O or S. 

DLA metallicities are found to be relatively low at high redshift,
some as low as $ [X/H] \sim -2.5$ in general, while some are substantially higher, up to $\sim$ -1 dex \citep{2014ApJ...782L..29R, 2018MNRAS.473.3559P, 2020MNRAS.491.1008P}. Our results indicate a diverse distribution of metallicity, with some DLAs being far more metal-enriched, including one DLA with a super-solar metallicity \citep{2023ApJ...954L..19H}. The host galaxies of such metal-rich DLAs appear to have experienced accelerated metal enrichment, suggestive of a burst of star formation within the first $\sim 1$ billion years of cosmic time. Over-density of [O III] emitters and galaxies with high star-formation rate are reported at redshift $z>6$. We speculate that the DLAs with high metallicity may potentially be the counterparts of the recently discovered high-redshift galaxies that show evidence of early bursts of star formation \citep{2025NatAs...9..280D}. 

\begin{figure}
    \centering
    \includegraphics[width=1\linewidth]{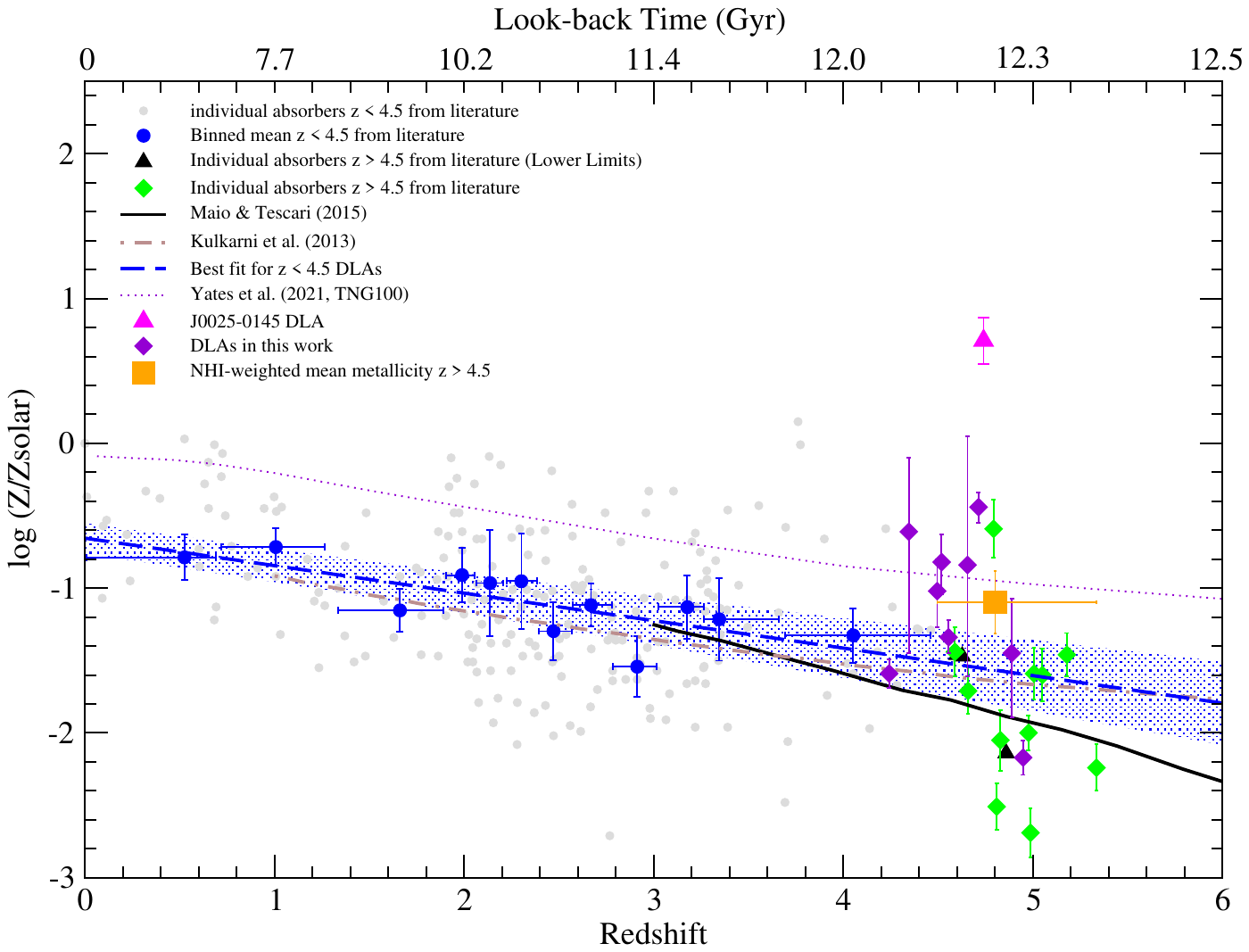}
    \caption{Metallicity-redshift relationship for DLAs from our survey as well as the literature. The small grey points show individual DLAs from the literature at $z < 4.5$, and the blue circles show $N_{\rm H I}$-weighted mean metallicities for the DLAs in various redshift bins. The purple diamonds and green diamonds show the DLAs at $z > 4.5$ from this work and the literature. The magenta triangle shows a super-solar metallicity DLA toward J0025-0145 reported in \citet{2023ApJ...954L..19H}. The orange square shows the binned $N_{\rm H I}$-weighted mean metallicity for the DLAs at $z > 4.5$, including all the measurements in this work and the literature except the supersolar metallicity DLA toward J0025-0145. DLAs at $z > 4.5$ show a wider diversity in metallicities compared to those at $z < 4.5$. The dashed blue line shows the best linear regression fit for the binned relation for DLAs at $z < 4.5$ given by $\mathrm{[M/H]} = -0.190\times z_{abs}-0.655$, and the shaded region in blue shows the $\pm 1 \sigma$ uncertainty in this best fit  for the $z < 4.5$ DLAs. All of the metallicity measurements shown are determined from undepleted elements (Zn, S, or O). The solid black, dot-dashed brown, and dotted purple curves show predicted metallicity-redshift relations from various models. Our findings suggest a relatively smooth DLA metallicity-redshift relation.}
    \label{fig:Zzmodel}
\end{figure}

Fig. \ref{fig:Zzmodel} shows the metallicity measurements for DLAs in our survey compared to metallicity measurements for other DLAs in the literature as well as predictions for the global metallicity-redshift relation in different models. All data points (both from our survey and the literature) are based on weakly depleted (volatile) elements. 
For DLAs at $z < 4.5$, the small grey circles show the measurements for the individual DLAs from the literature, and the blue circles show the binned $N_{\rm H I}$-weighted mean metallicities, with each bin consisting of 16-19 DLAs. For each bin, the $N_{\rm H I}$-weighted mean metallicity and the 1 $\sigma$ uncertainty (including measurement as well as sampling uncertainties) in the mean were calculated using survival analysis (which allows us to deal with a mixture of detections and limits), following the procedures described in \citet{2002ApJ...580..732K} and other past studies \citep[e.g.,][]{2005ApJ...618...68K, 2015ApJ...806...25S, 2020MNRAS.491.1008P}. For $z \ge 4.5$, we show the individual absorbers from our survey \citep[this paper and ][]{2023ApJ...954L..19H} and the literature \citep{2012ApJ...755...89R, 2014ApJ...782L..29R,2016ApJ...830..158M,2018MNRAS.473.3559P,2020MNRAS.491.1008P}. We do not include the $z \sim 6$ absorbers from \cite[e.g.,][]{2024A&A...687A.314S} because those were selected on the basis of not H~{\sc i}, but O~{\sc i}, and are thus not suitable for a statistical study of metallicity. Moreover, while H~{\sc i} column densities have been estimated for some of these absorbers, they are more uncertain due to the higher contamination from the IGM absorption at $z \sim 6$. Additionally, most of the $z \sim 6$ absorbers with H~{\sc i} estimates are proximate DLAs with $z_{abs} \approx z_{em}$), which makes them non-representative of the general DLA population.

The orange square in Fig. \ref{fig:Zzmodel} 
shows the $N_{\rm H I}$-weighted mean metallicity for all of the $z>4.5$ absorbers shown from our work and the literature, other than the extremely metal-rich absorber from \citet{2023ApJ...954L..19H}. 
(We note that nearly the same mean metallicity, within 0.04 dex, is obtained if the sub-DLAs at $z> 4.5$ are excluded.) The linear regression fit for the $N_{\rm H I}$-weighted mean metallicity of $z< 4.5$ DLAs is given by 
$$\mathrm{[M/H]} = (-0.190\pm0.045)\, z_{DLA}-(0.655\pm0.104),$$ which is shown with the blue dashed line (best-fit relation), 
and the shaded blue region ($\pm$ 1$\sigma$ uncertainty in the relation). There is no indication of a sudden drop in the $N_{\rm H I}$-weighted mean metallicity at $z > 4.5$. In fact, the $N_{\rm H I}$-weighted mean metallicity deduced from the $z> 4.5$ DLAs in our survey and the literature is slightly elevated (by 0.23 dex, i.e., $\sim$ 2$\sigma$) compared to the prediction from the $N_{\rm H I}$-weighted mean metallicity trend for $3.7 < z< 4.5$ DLAs. We note that this is not an artifact of sample selection, as our sample was selected based on the existence of strong H I Ly-$\alpha$ absorption features, rather than the strength of metal lines [see also \citep{2020MNRAS.491.1008P} for a detailed discussion of the lack of a selection bias toward metal-rich absorbers in their sample]. 

For comparison, we also show the metallicity-redshift relations predicted by models based on initial mass functions (IMFs) without \citep{2013ApJ...772...93K} and with \citep{2015MNRAS.453.3798M} Pop III stars, and results from IllustrisTNG-100 simulations \citep{2021MNRAS.508.3535Y}. The global mean metallicity (i.e., the $N_{\rm H I}$-weighted mean metallicity) of the absorbers at $z > 4.5$ lies within the range of predictions of these models: while it lies substantially above the predictions of the models of \citet{2013ApJ...772...93K} and \citet{2015MNRAS.453.3798M}, it is more consistent with the prediction from 
\citet{2021MNRAS.508.3535Y} (although we note that the latter over-predicts the 
metallicities at lower redshifts). 

\begin{figure}
    \centering
    \includegraphics[width=1\linewidth]{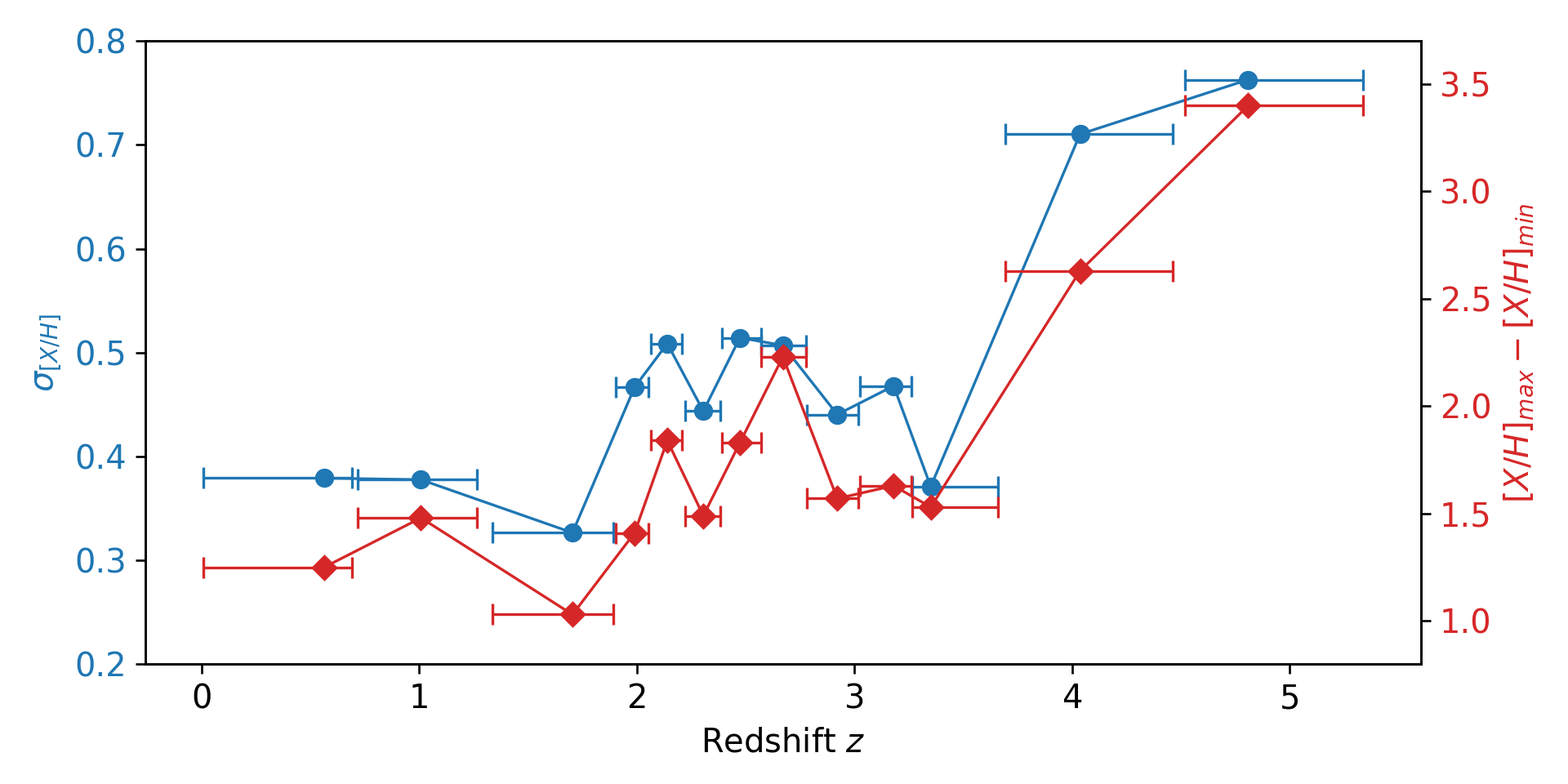}
    \caption{The scatter in DLA metallicities as a function of redshift. Blue circles denote 1 standard deviations in the metallicities  ($\sigma_{\mathrm{[X/H]}}$) of individual DLAs in the various redshift bins, and red diamonds denote the range of metallicities ($\mathrm{[X/H]_{\rm max} - \mathrm [X/H]_{\rm min}}$) for DLAs in various redshift bins. Both the metallicity range and the standard deviation show a increasing trend for the bins at redshift $3.7<z<4.5$ and $4.5<z<5.3$ compared to those at $z < 3.7$. }
    \label{fig:Zzmodel-scattering}
\end{figure}

The metallicity measurements at $z>4.5$ show a wide diversity, spanning $\sim$3 orders of magnitude. Fig. \ref{fig:Zzmodel-scattering} shows the scatter in the DLA metallicities in various redshift bins. The standard deviation of DLA metallicities ($\sigma_{\langle Z \rangle}$, shown as blue circles in Fig. \ref{fig:Zzmodel-scattering}) shows a substantial increase in the two highest redshift bins. A similar trend is also seen in the range of metallicity values (i.e., the difference between the maximum and minimum metallicities) in each bin. While the individual DLAs (shown as grey dots in 
Fig. \ref{fig:Zzmodel}) at $z < 3.7$ also show a substantial scatter in metallicity spanning $\sim$2 orders of magnitude, the scatter appears to be larger at $3.7<z<4.5$ and $4.5<z<5.3$, spanning $\sim$ 2.5-3.5 orders of magnitude. This diversity may result from different rates of chemical evolution in the DLA host galaxies due to differences in the star formation history and/or the initial mass function (IMF). 
Indeed, an increase in the metallicity dispersion of galaxies at higher redshift is also found in studies of cosmic chemical evolution based on EAGLE simulations  \citep[e.g.,][]{2021MNRAS.508.3535Y}.
Furthermore, as discussed in more detail in sections 4.3 and 4.4, the more metal-rich absorbers may also probe smaller impact parameters (i.e., regions closer to the galaxy centers). This may suggest a wider diversity of star formation histories or IMF variations at higher redshifts.
We also note that the dispersion in DLA metallicities shows a smaller peak at $2 < z < 3$ compared to DLAs at $z < 2$, potentially reflecting the effect of the much higher cosmic star formation activity at those epochs. 

Finally, we note that the wide diversity of metallicities at $z > 4.5$ is unlikely to be caused by the presence of some sub-DLAs in the $z > 4.5$ sample, since it persists even if the sub-DLAs are excluded. Likewise, the wide range of metallicities does not appear related to the presence of some proximate DLAs. Indeed, we do not find a significant difference between the metallicities of proximate DLAs and intervening DLAs within the limited sample of DLAs at $z> 4$. We plan to investigate the effects of the quasar proximity zone on high-redshift DLAs using a larger sample in a future paper.

\subsection{Are metals in DLAs at high redshift contributed from Pop III stars?}
The relative abundances of different elements in DLAs (defined in the standard way as [X/Y] = log$(N_{X}/N_{Y})$ - log$(X/Y)_{\odot}$ can potentially offer important constraints on the IMFs of the earlier generations of stars that enriched the gas in the DLA \citep[e.g.,][]{2013ApJ...772...93K,2014ApJ...787...64K}.

Past studies have investigated very metal-poor and extremely metal-poor DLAs at $ z_{abs} \sim 2-3$ and found their relative abundances to be similar to the signatures of very metal-poor stars, suggesting the metals in such metal-poor DLAs could potentially have formed from nucleosynthesis in Population III stars \citep{2011MNRAS.412.1047C,2011MNRAS.417.1534C,2019MNRAS.487.3363W,2022ApJ...929..158W}. For example, \cite{2022ApJ...929..158W} reported O-enhancement for two near-pristine DLAs at $z \sim 3.2$ and estimated the masses of the progenitor stars that could produce the metals to be (19-25) M$_{\odot}$. Our results for DLA/sub-DLAs at $z > 4$ show that O-enhancement also appears to be present in relatively metal-enriched DLAs at higher redshifts. In addition, we find S, another $\alpha$-element, to be enhanced in two of 10 DLAs in our survey. Given these unusual relative abundances, it is interesting to investigate whether the metals in high-$z$ DLAs reflect a leftover signature of enrichment by Pop III stars.

\begin{figure}
    \centering
    \includegraphics[width=1\linewidth]{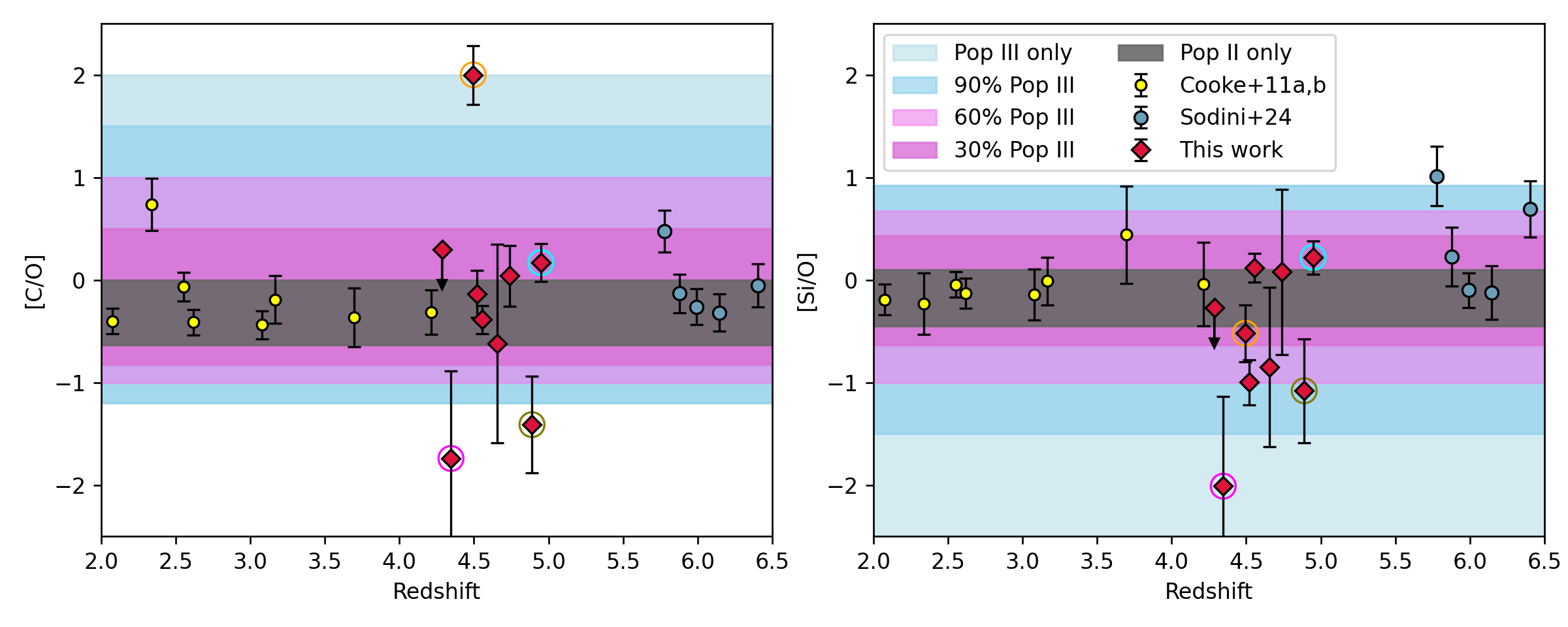}
    \caption{Comparison of the dust-depletion corrected abundances [C/O] and [Si/O] for DLAs from our study and the literature with the the maximum extent of [C/O] and [Si/O] in simulations from \cite{2023MNRAS.526.2620V} for different levels of enrichment from Pop III stars. A selected sample of very metal-poor DLAs from \cite{2011MNRAS.412.1047C,2011MNRAS.417.1534C} and references therein are shown as filled yellow circles. Proximate DLAs from \citep{2024A&A...687A.314S} shown as filled blue circles. Red diamonds  show DLAs from our survey, and diamonds enclosed in circles with  different colors are the DLAs that are consistent with a non-zero contribution from Pop III stars (J1129: orange; J1135 at $z_{DLA}=4.346$: magenta; J2202: cyan; J2328 at $z_{DLA}=4.889$: green).}
    \label{fig:PopIII-contribution}
\end{figure}

With this in mind, we now examine the [C/O] and [Si/O] ratios in DLAs. 
It is essential to correct for dust depletion before these ratios can be used to study nucleosynthetic signatures. We therefore correct the observed 
[C/O] and [Si/O] for dust depletion effects, following the procedure outlined in sec. \ref{sec:DTM}, with coefficients for C ($A2_{C} = 0.00$ and $B2_{C} = -0.10 \pm 0.10$) from \cite{2024A&A...681A..64K}. 
Fig. \ref{fig:PopIII-contribution} shows the dust-depletion corrected relative abundances of [C/O] and [Si/O] for the 9 DLAs in our survey for which [C/O] and [Si/O] can be constrained, along with the very metal-poor DLAs from \citet[][and references therein]{2011MNRAS.412.1047C, 2011MNRAS.417.1534C} and the proximate DLAs at high redshift from \cite{2024A&A...687A.314S}. We have corrected the [C/O] and [Si/O] ratios for all of these DLAs (from both our sample and the literature) for dust depletion. It is worth noting that the depletion correction raises the [Si/O] values compared to the observed (gas-phase) values, since 
Si is more depleted (refractory) compared to O. 

To investigate the potential signature of enrichment by Pop III stars, we also show in Fig. \ref{fig:PopIII-contribution} the maximum extent of [C/O] and [Si/O] predicted by the simulations of  \citet{2023MNRAS.526.2620V} for different levels of enrichment from Pop III stars. Despite the substantial  uncertainties in a few cases, 4 out of the 9 DLAs from our survey appear to be consistent with a non-zero contribution from Pop III stars. These 4 DLAs are shown with red diamonds enclosed in circles with different colors in Fig. \ref{fig:PopIII-contribution} (with 3 of these 4 DLAs being consistent with a $\gtrsim 30 \%$ contribution from Pop III stars). 
This suggests that the gas probed by at least some high-$z$ DLAs (with both low and high metallicities) may have been enriched by nucleosynthesis in Pop III stars, and that the signature of enrichment by Pop III stars can still remain at $z\sim4-5$. Our results corroborate a consistency with Pop III enrichment suggested by  \cite{2024A&A...687A.314S}, and more than triple the sample of high-$z$ absorbers with potential Pop III signatures in [C/O] and in [Si/O]. The very metal-poor DLAs from \citet[][and references therein]{2011MNRAS.412.1047C, 2011MNRAS.417.1534C}, which are at lower redshifts than the redshift range of our sample and the sample of \cite{2024A&A...687A.314S}, appear to show such Pop III signatures less commonly. 

We also note that the J1129 DLA (shown in Fig. \ref{fig:PopIII-contribution} with a red diamond enclosed in an  orange circle) appears to be a C-enhanced DLA, with a surprisingly high [C/O] relative abundance. Unfortunately, the [C/Fe] ratio could not be determined in this DLA, since the Fe abundance in the J1129 DLA is not available due to the very noisy Fe II $\lambda 1608$ line. Follow-up higher SNR observations covering Fe II absorption lines would help to further investigate the C-enhanced nature of this DLA. 

\subsection{$\alpha$-Element Enhancement and Element abundance patterns}

In Fig. \ref{fig:element-pattern}, we compare the element abundance patterns for the DLAs in our survey. We clearly see an $\alpha$-element-enhancement trend for O, Mg, Si and S. For example, the DLA in the sight line to J1135-1526 at $z_{DLA}=4.55538$ shows enhanced [O/Fe]=$+0.49$, [Si/Fe]=$+0.47$, and [S/Fe]=$+0.79$, and the DLA in the sight line to J1336-1830 at $z_{DLA}=4.518762$ shows enhanced [O/Fe]=$+1.96$ and [Si/Fe]=$+0.61$. The DLA in the sight line to J0025-0145 from our survey reported in an earlier paper \citep{2023ApJ...954L..19H} was also found to show [O/Fe] = +2.29 $\pm$ 0.05, [Si/Fe]= +1.02 $\pm$ 0.35, and [Mg/Fe] = +2.11$^{+0.27}_{-0.50}$ dex. 

{In our survey, 3 of the DLAs have both Si and Fe measurements, which enable  the dust depletion correction  following the method of \cite{2018A&A...611A..76D}, as outlined in sec. \ref{sec:DTM}. For comparison, we select metal-poor DLAs with $\mathrm{log} N_{HI} \gtrsim 20.0$ $\mathrm{cm}^{-2}$ in the literature \citep{2022ApJ...929..158W,2011MNRAS.417.1534C} and calculate their dust depletions of O and Fe following the same method. We then derive the depletion-corrected relative abundance ratios [$\alpha$/Fe]$_{\rm tot}$ based on O. We notice that the $[\alpha/Fe]$ ratios for some DLAs are lowered after the dust-depletion correction by as much as $\sim$1 dex. 3 of the DLAs in our survey show $\alpha$-enhancement after dust-depletion correction. This finding is consistent with what we find for the 21 metal-poor DLAs from the literature: 14 of these literature DLAs show lowering of the [$\alpha$/Fe] ratios after depletion correction but 6 of these 14 DLAs still show enhanced [$\alpha$/Fe] ratio after depletion correction. Our results demonstrate that dust depletion plays a significant role in determining relative abundances, and must be accounted for while investigating $\alpha$ enhancement and other relative abundances in DLAs over a wide range of redshifts and spanning both low as well as high metallicities. 

The $\alpha$-element enhancement for DLAs along the sight lines to the afterglows of gamma-ray bursts (GRBs) was studied by \cite{2007ApJ...666..267P}, who noted that both [$\alpha$/Fe] and [Zn/Fe] ratios for GRB DLAs are higher than those for DLAs along quasar sight lines (QSO DLAs). This difference is attributed to a difference in the galactocentric radius for typical GRB DLAs and QSO DLAs, as GRB DLAs preferentially trace denser and more metal-enriched ISM in regions closer to the galaxy centers compared to QSO DLAs. 

In our survey, we find that DLAs with higher metallicity show stronger [$\alpha$/Fe] enhancement as shown in Fig. \ref{fig:element-pattern} and Fig. \ref{fig:alpha-enhancement}. Similar to the comparison between GRB DLAs and QSO DLAs discussed above, the comparison of highly scattered DLA metallicities at high redshift may suggest that the sight lines probed by the metal-rich QSO DLAs pass through the inner regions of the host galaxies, compared to the sight lines probed by the metal-poor DLAs. Interestingly, [$\alpha$/Fe] enhancement may also be more prominent in the inner regions of host galaxies, indicating that massive stars may be more common at lower impact parameter than at higher impact parameter relative to galaxies at high redshift.

\subsection{Dust-to-Metal ratios}

Fig. \ref{fig:Fstar_Z} and Fig. \ref{fig:DTM} compare our estimates of $F_*$ and DTM for the DLAs in our survey and other DLAs from the literature, including both DLAs at $z > 4.5$ \citep{2016ApJ...830..158M, 2018MNRAS.473.3559P, 2020MNRAS.491.1008P, 2019ApJ...885...59B} as well as metal-poor DLAs \citep{2022ApJ...929..158W}, thus covering a wide range of redshifts, observed [Fe/H], and DTM$_{Fe}$.

We test the correlation of $F_*$ versus intrinsic metallicity ${\rm Z}_{\rm instrinsic}$, and DTM$_{Fe}$ versus [Fe/H]$_{\rm tot}$ using Spearman rank-order correlation test. The Spearman rank-order correlation coefficient and the corresponding probability are $r_{S}=0.546$ and $p=3.875\times 10^{-4}$ for $F_*$ versus ${\rm Z}_{\rm instrinsic}$. The correlation test for DTM$_{Fe}$ and [Fe/H]$_{\rm tot}$ gives $r_{S}=0.719$ and $p<10^{-10}$ for the full sample of DLAs with a detection of Fe depletion (not an upper limit). This correlation is dominated by the higher metallicity DLAs, although even the subset of DLAs with [Fe/H]$_{\rm tot}<-1$ show a strong correlation between DTM$_{Fe}$ and [Fe/H]$_{\rm tot}$ ($r_{S}=0.416$, $p=2.94 \times10^{-6}$). In general, absorbers with higher metallicity appear to have higher dust depletion (higher $F_*$ values and DTM$_{Fe}$ ratios).

\cite{2019A&A...623A..43B} reported a higher fraction of molecular H$_2$ detections for GRB DLAs compared to QSO DLAs, which may indicate the absorbing gas probed by GRB DLAs lies closer to the host galaxy center, where the higher gas pressure makes the conversion of H~{\sc i} to H$_2$ more readily possible \cite[e.g.,][]{2006ApJ...650..933B}. More interestingly, H$_2$-bearing GRB DLAs reported in \cite{2019A&A...623A..43B} are found to be associated with significant dust content with DTM$>0.4$. While the H$_{2}$ content of the high-$z$ DLAs is not yet known, the correlation of DTM$_{Fe}$ and [Fe/H]$_{\rm tot}$ seen in Fig. 6 suggests that metal-rich intervening DLAs may also potentially trace inner regions of the absorbing host galaxies, where denser, more metal-enriched and dustier gas is expected. The higher dust content of metal-enriched DLAs may trace higher levels of molecular content and higher star formation rates. If high star formation rates existed in the first $\sim$ 1 billion years after the Big Bang, they could have contributed to the metal enrichment observed in these high-$z$ DLAs.

\begin{figure}
    \centering
    \includegraphics[width=1\linewidth]{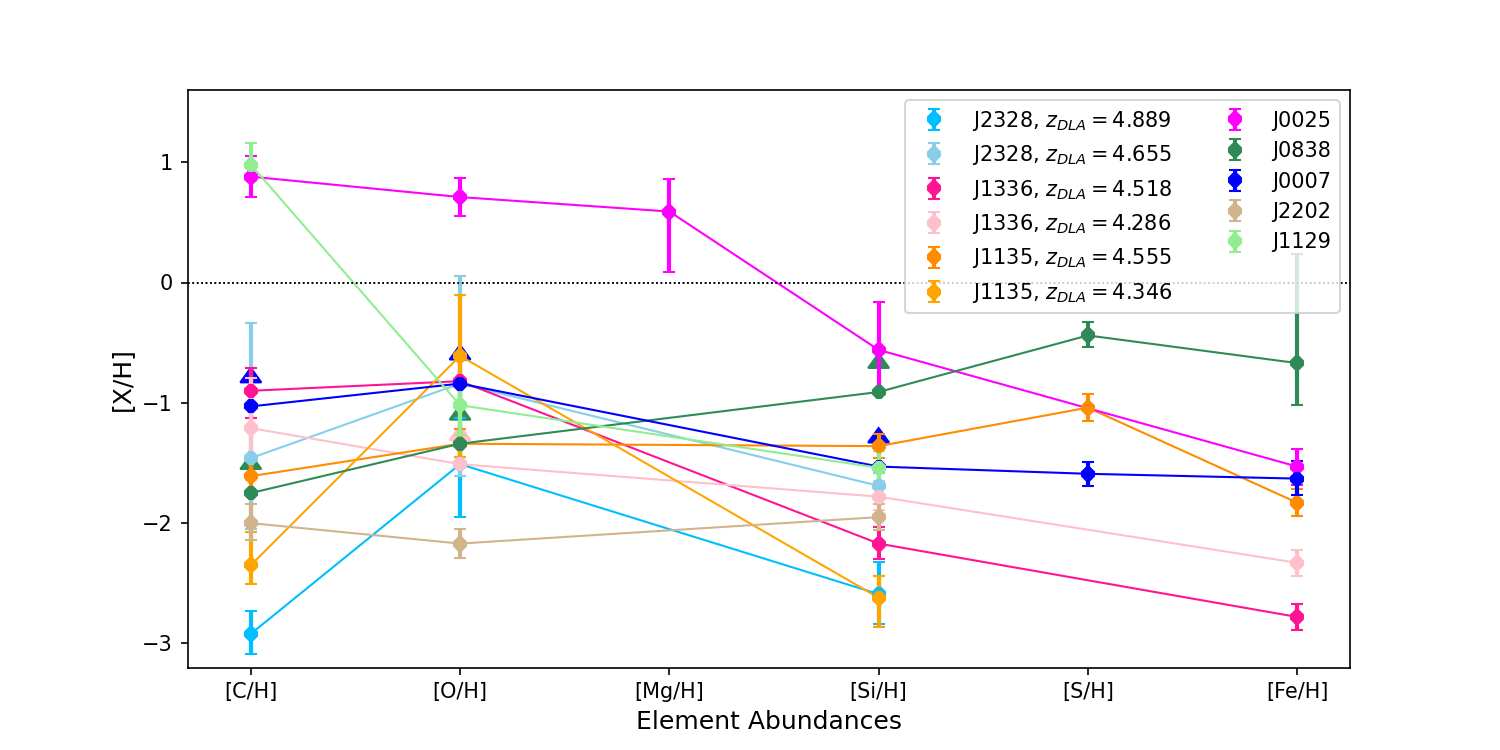}
    \caption{The chemical abundance patterns are overplotted in different colors for the different DLAs from our survey presented in this work and \citet{2023ApJ...954L..19H}. The chemical enrichment [X/H] is plotted in the order of X$=$C, O, Mg, S, Si and Fe on the y-axis. Comparison of different elements reveals enhanced observed [$\alpha$/Fe] ratios for O, Mg, Si and S for several DLAs.}
    \label{fig:element-pattern}
\end{figure}

\begin{figure}
    \centering
    \includegraphics[width=\linewidth]{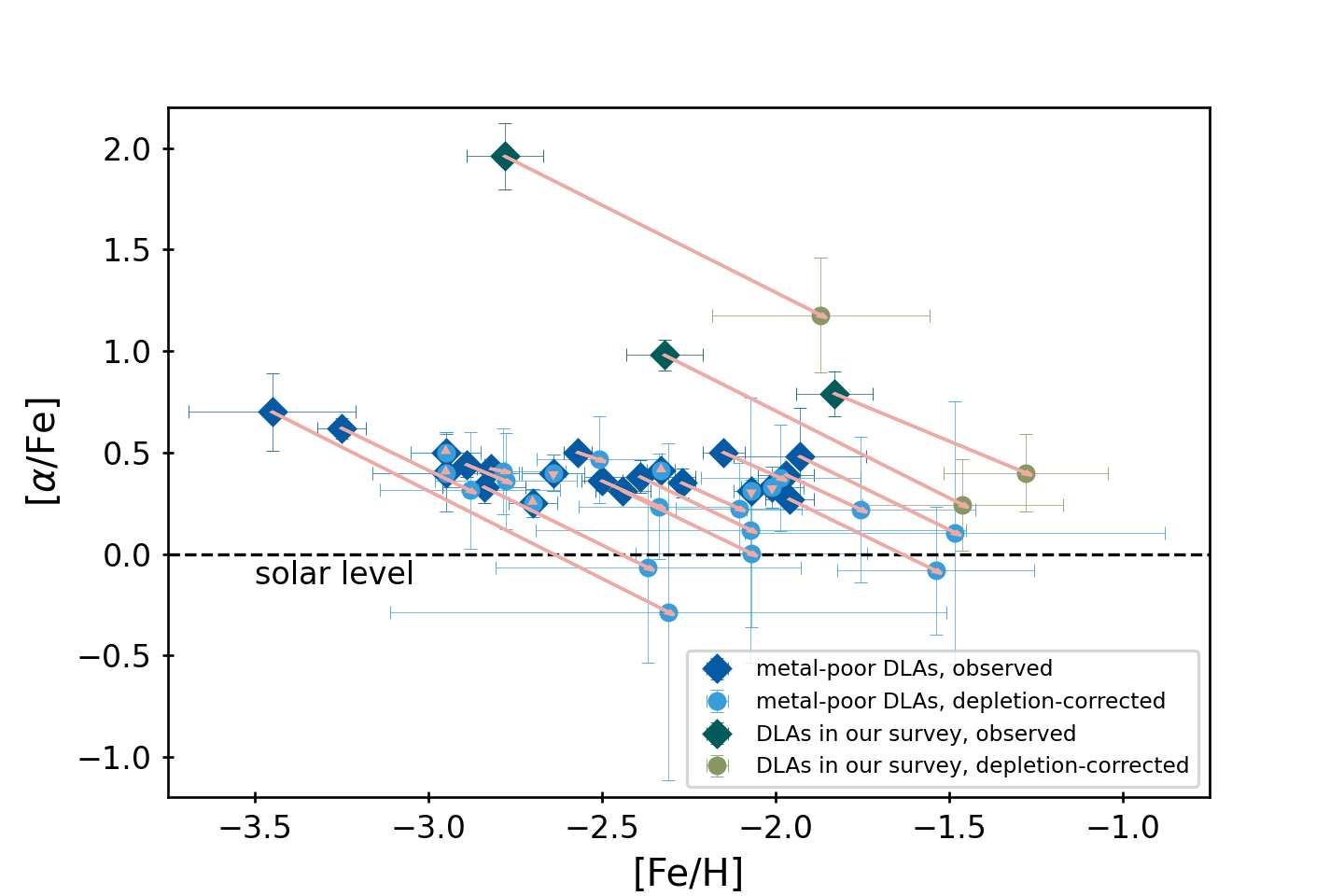}  
    \caption{[$\alpha$/Fe] ratios for DLAs in our survey (green) and metal-poor DLAs (blue) from the literature  \citep[][and references therein]{2022ApJ...929..158W, 2011MNRAS.417.1534C}. The [$\alpha$/Fe] ratios from the literature are based on O, and the [$\alpha$/Fe] ratios in our survey are based on O or S (when O is not well-constrained due to line saturation). Blue and green diamonds denote observed values from the literature and our survey, while blue and green circles denote dust-depletion corrected values for the corresponding DLAs, where both the $\alpha$-element and Fe are corrected as described in sec. \ref{sec:DTM}. Pink lines join the observed and dust depletion-corrected values. This figure illustrates that dust-depletion correction can decrease the enhancement of observed [$\alpha$/Fe] ratios. For 3 DLAs in our survey, the [$\alpha$/Fe] enhancement is still evident after applying the dust-depletion correction.}
    \label{fig:alpha-enhancement}
\end{figure}

\begin{figure}
    \centering
    \includegraphics[width=1\linewidth]{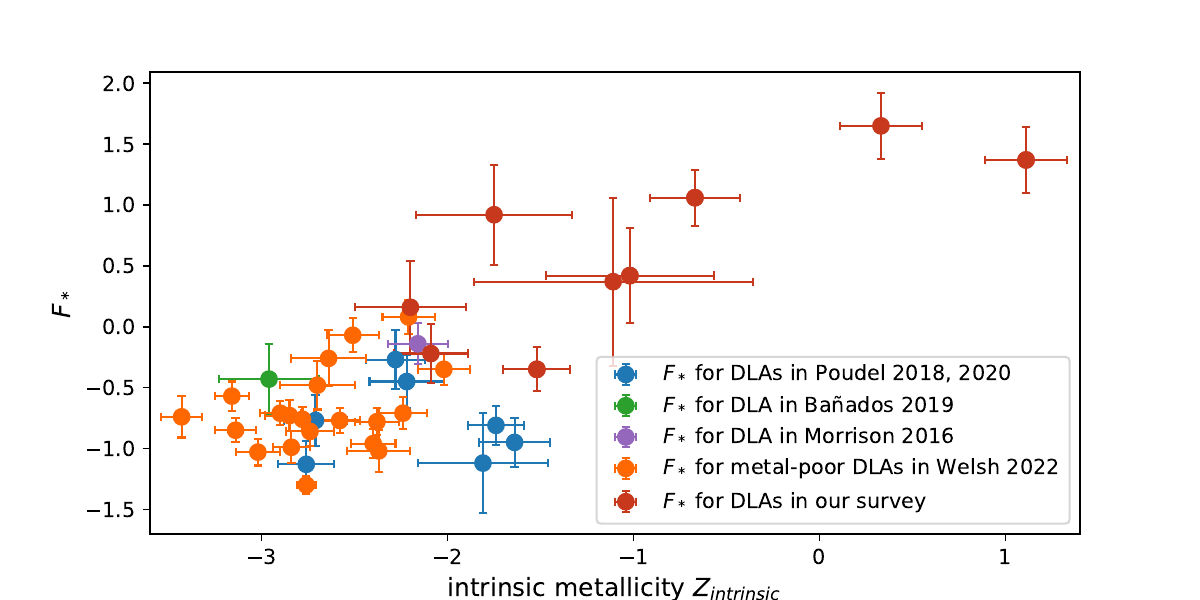}
    \caption{The line-of-sight depletion factor $F_{*}$ vs. the depletion-corrected intrinsic metallicity, estimated by applying the Jenkins (2009) method for our absorbers and those from the literature. The depletion-corrected intrinsic metallicity shows a positive correlation with $F^*$, and the metal-rich DLAs are estimated to be much dustier than metal-poor DLAs ($\Delta F^* > 2$).}
    \label{fig:Fstar_Z}
\end{figure}

\begin{figure}
    \centering
    \includegraphics[width=1\linewidth]{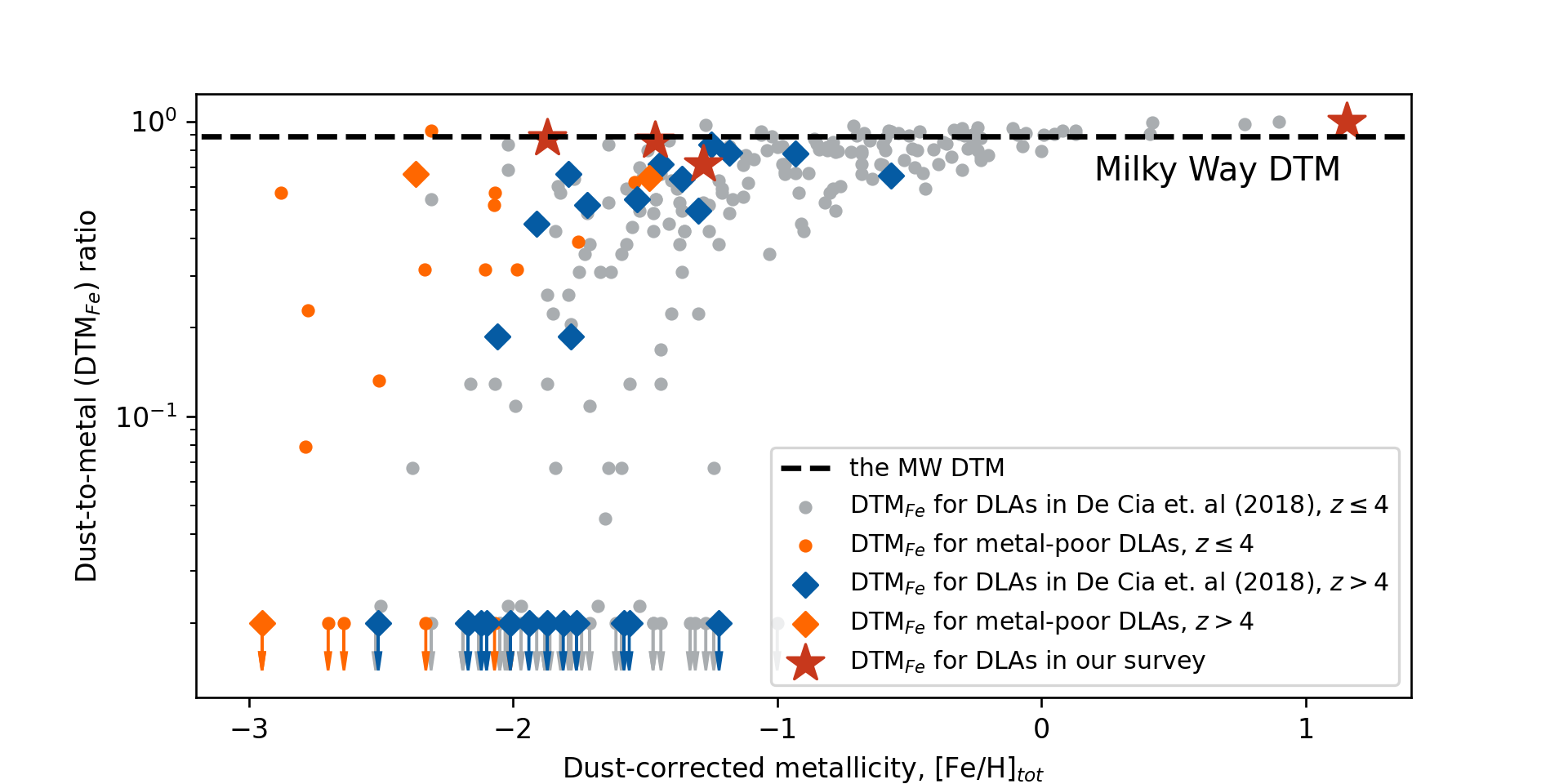}
    \caption{Dust-to-metal ratios vs. dust depletion-corrected metallicities, estimated by applying the methods of \citet{2016A&A...596A..97D, 2018A&A...611A..76D} to DLAs in our survey (shown as dark red stars), and other DLAs from the literature at lower redshifts (light grey and orange dots) and at similar redshifts (dark blue and orange diamonds). Data points in orange are selected metal-poor DLAs from \citet{2022ApJ...929..158W}, \citet{2011MNRAS.417.1534C}, and references therein. Data points in grey are from \citet{2018A&A...611A..76D} and references therein. Overall, the full DLA sample shows a positive correlation of DTM$_{Fe}$ ratios with the depletion-corrected [Fe/H]$_{\rm tot}$.}
    \label{fig:DTM}
\end{figure}

\section{Conclusions}
We report the discovery of 10 DLAs/sub-DLAs at redshift $4.2 < z < 5.0$, and the measurements of the abundances of various elements including C, O, Si, S, and Fe in these absorbers. Our main conclusions are as follows:

1. Our survey shows DLAs at $z \ge 4$ to be generally metal-poor ($<$0.1 solar) as found in past studies; however, our observations indicate a wide diversity of metallicities, with a substantial fraction of relatively metal-enriched DLAs ($>$ 0.1 solar) including a DLA at $z \sim 4.7$ that appears to have a super-solar metallicity. DLAs at $z>4.5$ appear to have a wider diversity in metallicity compared to those at $z<4.5$. Combining our sample with the literature, we find that the DLA metallicity appears to show a relatively smooth evolution with redshift out to $z \sim 5.3$, with a tentative ($\sim 2\sigma$) indication of a slight rise in metallicity at $z > 4.5$.

2. We find a larger metallicity dispersion among DLAs at $3.7 < z < 4.5$ and $4.5 < z < 5.3$ compared to DLAs at $z < 3.7$, as also expected in some cosmic chemical evolution models, potentially suggesting a wider diversity in star formation histories or IMFs of the higher redshift DLA host galaxies.

3. Comparing the depletion-corrected relative abundances [C/O] and [Si/O] with the maximum extent of these ratios predicted by models including Pop III stars, we find that a significant fraction of DLAs at $z>4$ show potential signatures of enrichment from Pop III stars, and that several of these DLAs are consistent with a $\gtrsim30\%$ contribution of metals from Pop III stars. This suggests that signatures of enrichment by Pop III stars can still remain at $z \sim 4-5$. 

4. We find several cases with the observed O abundance higher than that of Fe. We also find two detections of S (which, like O, is also nearly undepleted on dust grains), and find S to be enhanced compared to Fe. Thus, besides C-enhancement and O-enhancement, S-enhancement can also be present. 

5. Comparing the volatile and refractory elements, we determined the line-of-sight dust depletion factor $F_{*}$ and the depletion-corrected metallicities for our DLAs, and thereby determined the dust-to-metals ratios based on Fe 
for DLAs in our survey and the literature. Combining our sample and the literature, we find both $F_{*}$ and DTM$_{Fe}$ to correlate positively with the 
depletion-corrected metallicity. This indicates that galaxies hosting these relatively metal-enriched absorbers are dusty. 

6. We also investigated the effect of dust depletion on the $\alpha$-enhancement relative to Fe in DLAs. We find that the $\alpha$-enhancement is lowered after the dust depletion corrections. Despite this, all 3 DLAs in our survey that have undergone dust depletion corrections and those from the literature (6 out of 21) still show significant $\alpha$-enhancement even after the dust depletion correction. This intrinsic $\alpha$-enhancement in these DLAs suggests that their absorbing gas may (partly) have been enriched from core-collapse supernovae. Our results also suggest that the DLAs showing high metallicities and high dust depletion factors may be probing regions closer to the centers of the host galaxies, and may have experienced accelerated chemical enrichment due to higher star formation rates at earlier epochs.

\begin{acknowledgments}

We thank an anonymous referee for insightful comments that have helped to improve this paper.
J.H. and V.P.K. acknowledge partial support from the National Science Foundation grants AST/2007538 and AST/2009811 (PI: Kulkarni). 
V.P.K. gratefully acknowledges additional support from NASA grant 80NSSC20K0887 (PI: Kulkarni). 
S.P. would like to express gratitude for the support received from the ESOChile joint committee grant. 
N.T. and S.L. acknowledge support by FONDECYT grant 1231187. 
This paper incorporates data collected through the programs CN2021B-57, CN2022A-61 (PI: Poudel), and NOIRLab programs 2023A-462536, 2023B-190445 (PI: Huyan) using the MIKE spectrograph on the Magellan Clay telescope, situated at Las Campanas
Observatory, Chile. We thank the Las Campanas Observatory
staff for assistance in conducting the observations.
\end{acknowledgments}

\noindent{Facilities: Magellan:Clay}

\noindent{Software: Astropy (The Astropy Collaboration et al. 2013, 2018), VoigtFit (Krogager 2018), Linetools (Prochaska
et al. 2017).}

\newpage

\bibliographystyle{aasjournal}
\bibliography{paper_refs}

\begin{thebibliography}{}
\expandafter\ifx\csname natexlab\endcsname\relax\def\natexlab#1{#1}\fi
\providecommand{\url}[1]{\href{#1}{#1}}
\providecommand{\dodoi}[1]{doi:~\href{http://doi.org/#1}{\nolinkurl{#1}}}
\providecommand{\doeprint}[1]{\href{http://ascl.net/#1}{\nolinkurl{http://ascl.net/#1}}}
\providecommand{\doarXiv}[1]{\href{https://arxiv.org/abs/#1}{\nolinkurl{https://arxiv.org/abs/#1}}}

\bibitem[{{Astropy Collaboration} {et~al.}(2013){Astropy Collaboration},
  {Robitaille}, {Tollerud}, {Greenfield}, {Droettboom}, {Bray}, {Aldcroft},
  {Davis}, {Ginsburg}, {Price-Whelan}, {Kerzendorf}, {Conley}, {Crighton},
  {Barbary}, {Muna}, {Ferguson}, {Grollier}, {Parikh}, {Nair}, {Unther},
  {Deil}, {Woillez}, {Conseil}, {Kramer}, {Turner}, {Singer}, {Fox}, {Weaver},
  {Zabalza}, {Edwards}, {Azalee Bostroem}, {Burke}, {Casey}, {Crawford},
  {Dencheva}, {Ely}, {Jenness}, {Labrie}, {Lim}, {Pierfederici}, {Pontzen},
  {Ptak}, {Refsdal}, {Servillat}, \& {Streicher}}]{2013A&A...558A..33A}
{Astropy Collaboration}, {Robitaille}, T.~P., {Tollerud}, E.~J., {et~al.} 2013,
  \aap, 558, A33, \dodoi{10.1051/0004-6361/201322068}

\bibitem[{{Astropy Collaboration} {et~al.}(2018){Astropy Collaboration},
  {Price-Whelan}, {Sip{\H{o}}cz}, {G{\"u}nther}, {Lim}, {Crawford}, {Conseil},
  {Shupe}, {Craig}, {Dencheva}, {Ginsburg}, {VanderPlas}, {Bradley},
  {P{\'e}rez-Su{\'a}rez}, {de Val-Borro}, {Aldcroft}, {Cruz}, {Robitaille},
  {Tollerud}, {Ardelean}, {Babej}, {Bach}, {Bachetti}, {Bakanov}, {Bamford},
  {Barentsen}, {Barmby}, {Baumbach}, {Berry}, {Biscani}, {Boquien}, {Bostroem},
  {Bouma}, {Brammer}, {Bray}, {Breytenbach}, {Buddelmeijer}, {Burke},
  {Calderone}, {Cano Rodr{\'\i}guez}, {Cara}, {Cardoso}, {Cheedella}, {Copin},
  {Corrales}, {Crichton}, {D'Avella}, {Deil}, {Depagne}, {Dietrich}, {Donath},
  {Droettboom}, {Earl}, {Erben}, {Fabbro}, {Ferreira}, {Finethy}, {Fox},
  {Garrison}, {Gibbons}, {Goldstein}, {Gommers}, {Greco}, {Greenfield},
  {Groener}, {Grollier}, {Hagen}, {Hirst}, {Homeier}, {Horton}, {Hosseinzadeh},
  {Hu}, {Hunkeler}, {Ivezi{\'c}}, {Jain}, {Jenness}, {Kanarek}, {Kendrew},
  {Kern}, {Kerzendorf}, {Khvalko}, {King}, {Kirkby}, {Kulkarni}, {Kumar},
  {Lee}, {Lenz}, {Littlefair}, {Ma}, {Macleod}, {Mastropietro}, {McCully},
  {Montagnac}, {Morris}, {Mueller}, {Mumford}, {Muna}, {Murphy}, {Nelson},
  {Nguyen}, {Ninan}, {N{\"o}the}, {Ogaz}, {Oh}, {Parejko}, {Parley}, {Pascual},
  {Patil}, {Patil}, {Plunkett}, {Prochaska}, {Rastogi}, {Reddy Janga},
  {Sabater}, {Sakurikar}, {Seifert}, {Sherbert}, {Sherwood-Taylor}, {Shih},
  {Sick}, {Silbiger}, {Singanamalla}, {Singer}, {Sladen}, {Sooley},
  {Sornarajah}, {Streicher}, {Teuben}, {Thomas}, {Tremblay}, {Turner},
  {Terr{\'o}n}, {van Kerkwijk}, {de la Vega}, {Watkins}, {Weaver}, {Whitmore},
  {Woillez}, {Zabalza}, \& {Astropy Contributors}}]{2018AJ....156..123A}
{Astropy Collaboration}, {Price-Whelan}, A.~M., {Sip{\H{o}}cz}, B.~M., {et~al.}
  2018, \aj, 156, 123, \dodoi{10.3847/1538-3881/aabc4f}

\bibitem[{{Ba{\~n}ados} {et~al.}(2019){Ba{\~n}ados}, {Rauch}, {Decarli},
  {Farina}, {Hennawi}, {Mazzucchelli}, {Venemans}, {Walter}, {Simcoe},
  {Prochaska}, {Cooper}, {Davies}, \& {Chen}}]{2019ApJ...885...59B}
{Ba{\~n}ados}, E., {Rauch}, M., {Decarli}, R., {et~al.} 2019, \apj, 885, 59,
  \dodoi{10.3847/1538-4357/ab4129}

\bibitem[{{Becker} {et~al.}(2011){Becker}, {Sargent}, {Rauch}, \&
  {Calverley}}]{2011ApJ...735...93B}
{Becker}, G.~D., {Sargent}, W. L.~W., {Rauch}, M., \& {Calverley}, A.~P. 2011,
  \apj, 735, 93, \dodoi{10.1088/0004-637X/735/2/93}

\bibitem[{{Becker} {et~al.}(2006){Becker}, {Sargent}, {Rauch}, \&
  {Simcoe}}]{2006ApJ...640...69B}
{Becker}, G.~D., {Sargent}, W. L.~W., {Rauch}, M., \& {Simcoe}, R.~A. 2006,
  \apj, 640, 69, \dodoi{10.1086/500079}

\bibitem[{{Berg} {et~al.}(2016){Berg}, {Ellison}, {S{\'a}nchez-Ram{\'\i}rez},
  {Prochaska}, {Lopez}, {D'Odorico}, {Becker}, {Christensen}, {Cupani},
  {Denney}, \& {Worseck}}]{2016MNRAS.463.3021B}
{Berg}, T.~A.~M., {Ellison}, S.~L., {S{\'a}nchez-Ram{\'\i}rez}, R., {et~al.}
  2016, \mnras, 463, 3021, \dodoi{10.1093/mnras/stw2232}

\bibitem[{{Blitz} \& {Rosolowsky}(2006)}]{2006ApJ...650..933B}
{Blitz}, L., \& {Rosolowsky}, E. 2006, \apj, 650, 933, \dodoi{10.1086/505417}

\bibitem[{{Bolmer} {et~al.}(2019){Bolmer}, {Ledoux}, {Wiseman}, {De Cia},
  {Selsing}, {Schady}, {Greiner}, {Savaglio}, {Burgess}, {D'Elia}, {Fynbo},
  {Goldoni}, {Hartmann}, {Heintz}, {Jakobsson}, {Japelj}, {Kaper}, {Tanvir},
  {Vreeswijk}, \& {Zafar}}]{2019A&A...623A..43B}
{Bolmer}, J., {Ledoux}, C., {Wiseman}, P., {et~al.} 2019, \aap, 623, A43,
  \dodoi{10.1051/0004-6361/201834422}

\bibitem[{{Bunker} {et~al.}(2023){Bunker}, {Saxena}, {Cameron}, {Willott},
  {Curtis-Lake}, {Jakobsen}, {Carniani}, {Smit}, {Maiolino}, {Witstok},
  {Curti}, {D'Eugenio}, {Jones}, {Ferruit}, {Arribas}, {Charlot}, {Chevallard},
  {Giardino}, {de Graaff}, {Looser}, {L{\"u}tzgendorf}, {Maseda}, {Rawle},
  {Rix}, {Del Pino}, {Alberts}, {Egami}, {Eisenstein}, {Endsley}, {Hainline},
  {Hausen}, {Johnson}, {Rieke}, {Rieke}, {Robertson}, {Shivaei}, {Stark},
  {Sun}, {Tacchella}, {Tang}, {Williams}, {Willmer}, {Baker}, {Baum},
  {Bhatawdekar}, {Bowler}, {Boyett}, {Chen}, {Circosta}, {Helton}, {Ji},
  {Kumari}, {Lyu}, {Nelson}, {Parlanti}, {Perna}, {Sandles}, {Scholtz},
  {Suess}, {Topping}, {{\"U}bler}, {Wallace}, \&
  {Whitler}}]{2023A&A...677A..88B}
{Bunker}, A.~J., {Saxena}, A., {Cameron}, A.~J., {et~al.} 2023, \aap, 677, A88,
  \dodoi{10.1051/0004-6361/202346159}

\bibitem[{{Carniani} {et~al.}(2024){Carniani}, {Hainline}, {D'Eugenio},
  {Eisenstein}, {Jakobsen}, {Witstok}, {Johnson}, {Chevallard}, {Maiolino},
  {Helton}, {Willott}, {Robertson}, {Alberts}, {Arribas}, {Baker},
  {Bhatawdekar}, {Boyett}, {Bunker}, {Cameron}, {Cargile}, {Charlot}, {Curti},
  {Curtis-Lake}, {Egami}, {Giardino}, {Isaak}, {Ji}, {Jones}, {Kumari},
  {Maseda}, {Parlanti}, {P{\'e}rez-Gonz{\'a}lez}, {Rawle}, {Rieke}, {Rieke},
  {Del Pino}, {Saxena}, {Scholtz}, {Smit}, {Sun}, {Tacchella}, {{\"U}bler},
  {Venturi}, {Williams}, \& {Willmer}}]{2024Natur.633..318C}
{Carniani}, S., {Hainline}, K., {D'Eugenio}, F., {et~al.} 2024, \nat, 633, 318,
  \dodoi{10.1038/s41586-024-07860-9}

\bibitem[{{Chatzikos} {et~al.}(2023){Chatzikos}, {Bianchi}, {Camilloni},
  {Chakraborty}, {Gunasekera}, {Guzm{\'a}n}, {Milby}, {Sarkar}, {Shaw}, {van
  Hoof}, \& {Ferland}}]{2023RMxAA..59..327C}
{Chatzikos}, M., {Bianchi}, S., {Camilloni}, F., {et~al.} 2023, \rmxaa, 59,
  327, \dodoi{10.22201/ia.01851101p.2023.59.02.12}

\bibitem[{{Cooke} {et~al.}(2011{\natexlab{a}}){Cooke}, {Pettini}, {Steidel},
  {Rudie}, \& {Jorgenson}}]{2011MNRAS.412.1047C}
{Cooke}, R., {Pettini}, M., {Steidel}, C.~C., {Rudie}, G.~C., \& {Jorgenson},
  R.~A. 2011{\natexlab{a}}, \mnras, 412, 1047,
  \dodoi{10.1111/j.1365-2966.2010.17966.x}

\bibitem[{{Cooke} {et~al.}(2011{\natexlab{b}}){Cooke}, {Pettini}, {Steidel},
  {Rudie}, \& {Nissen}}]{2011MNRAS.417.1534C}
{Cooke}, R., {Pettini}, M., {Steidel}, C.~C., {Rudie}, G.~C., \& {Nissen},
  P.~E. 2011{\natexlab{b}}, \mnras, 417, 1534,
  \dodoi{10.1111/j.1365-2966.2011.19365.x}

\bibitem[{{Curtis-Lake} {et~al.}(2023){Curtis-Lake}, {Carniani}, {Cameron},
  {Charlot}, {Jakobsen}, {Maiolino}, {Bunker}, {Witstok}, {Smit}, {Chevallard},
  {Willott}, {Ferruit}, {Arribas}, {Bonaventura}, {Curti}, {D'Eugenio},
  {Franx}, {Giardino}, {Looser}, {L{\"u}tzgendorf}, {Maseda}, {Rawle}, {Rix},
  {Rodr{\'\i}guez del Pino}, {{\"U}bler}, {Sirianni}, {Dressler}, {Egami},
  {Eisenstein}, {Endsley}, {Hainline}, {Hausen}, {Johnson}, {Rieke},
  {Robertson}, {Shivaei}, {Stark}, {Tacchella}, {Williams}, {Willmer},
  {Bhatawdekar}, {Bowler}, {Boyett}, {Chen}, {de Graaff}, {Helton}, {Hviding},
  {Jones}, {Kumari}, {Lyu}, {Nelson}, {Perna}, {Sandles}, {Saxena}, {Suess},
  {Sun}, {Topping}, {Wallace}, \& {Whitler}}]{2023NatAs...7..622C}
{Curtis-Lake}, E., {Carniani}, S., {Cameron}, A., {et~al.} 2023, Nature
  Astronomy, 7, 622, \dodoi{10.1038/s41550-023-01918-w}

\bibitem[{{De Cia} {et~al.}(2016){De Cia}, {Ledoux}, {Mattsson}, {Petitjean},
  {Srianand}, {Gavignaud}, \& {Jenkins}}]{2016A&A...596A..97D}
{De Cia}, A., {Ledoux}, C., {Mattsson}, L., {et~al.} 2016, \aap, 596, A97,
  \dodoi{10.1051/0004-6361/201527895}

\bibitem[{{De Cia} {et~al.}(2018){De Cia}, {Ledoux}, {Petitjean}, \&
  {Savaglio}}]{2018A&A...611A..76D}
{De Cia}, A., {Ledoux}, C., {Petitjean}, P., \& {Savaglio}, S. 2018, \aap, 611,
  A76, \dodoi{10.1051/0004-6361/201731970}

\bibitem[{{de Graaff} {et~al.}(2025){de Graaff}, {Setton}, {Brammer}, {Cutler},
  {Suess}, {Labb{\'e}}, {Leja}, {Weibel}, {Maseda}, {Whitaker}, {Bezanson},
  {Boogaard}, {Cleri}, {De Lucia}, {Franx}, {Greene}, {Hirschmann}, {Matthee},
  {McConachie}, {Naidu}, {Oesch}, {Price}, {Rix}, {Valentino}, {Wang}, \&
  {Williams}}]{2025NatAs...9..280D}
{de Graaff}, A., {Setton}, D.~J., {Brammer}, G., {et~al.} 2025, Nature
  Astronomy, 9, 280, \dodoi{10.1038/s41550-024-02424-3}

\bibitem[{{Deepak} {et~al.}(2024){Deepak}, {Howk}, {Lehner}, \&
  {P{\'e}roux}}]{2024arXiv241119465D}
{Deepak}, S., {Howk}, J.~C., {Lehner}, N., \& {P{\'e}roux}, C. 2024, arXiv
  e-prints, arXiv:2411.19465, \dodoi{10.48550/arXiv.2411.19465}

\bibitem[{{D'Odorico} {et~al.}(2023){D'Odorico}, {Ba{\~n}ados}, {Becker},
  {Bischetti}, {Bosman}, {Cupani}, {Davies}, {Farina}, {Ferrara}, {Feruglio},
  {Mazzucchelli}, {Ryan-Weber}, {Schindler}, {Sodini}, {Venemans}, {Walter},
  {Chen}, {Lai}, {Zhu}, {Bian}, {Campo}, {Carniani}, {Cristiani}, {Davies},
  {Decarli}, {Drake}, {Eilers}, {Fan}, {Gaikwad}, {Gallerani}, {Greig},
  {Haehnelt}, {Hennawi}, {Keating}, {Kulkarni}, {Mesinger}, {Meyer},
  {Neeleman}, {Onoue}, {Pallottini}, {Qin}, {Rojas-Ruiz}, {Satyavolu},
  {Sebastian}, {Tripodi}, {Wang}, {Wolfson}, {Yang}, \&
  {Zanchettin}}]{2023MNRAS.523.1399D}
{D'Odorico}, V., {Ba{\~n}ados}, E., {Becker}, G.~D., {et~al.} 2023, \mnras,
  523, 1399, \dodoi{10.1093/mnras/stad1468}

\bibitem[{{Ferland} {et~al.}(2013){Ferland}, {Porter}, {van Hoof}, {Williams},
  {Abel}, {Lykins}, {Shaw}, {Henney}, \& {Stancil}}]{2013RMxAA..49..137F}
{Ferland}, G.~J., {Porter}, R.~L., {van Hoof}, P.~A.~M., {et~al.} 2013, \rmxaa,
  49, 137.
\newblock \doarXiv{1302.4485}

\bibitem[{{Helton} {et~al.}(2024){Helton}, {Sun}, {Woodrum}, {Hainline},
  {Willmer}, {Rieke}, {Rieke}, {Tacchella}, {Robertson}, {Johnson}, {Alberts},
  {Eisenstein}, {Hausen}, {Bonaventura}, {Bunker}, {Charlot}, {Curti},
  {Curtis-Lake}, {Looser}, {Maiolino}, {Willott}, {Witstok}, {Boyett}, {Chen},
  {Egami}, {Endsley}, {Hviding}, {Jaffe}, {Ji}, {Lyu}, \&
  {Sandles}}]{2024ApJ...962..124H}
{Helton}, J.~M., {Sun}, F., {Woodrum}, C., {et~al.} 2024, \apj, 962, 124,
  \dodoi{10.3847/1538-4357/ad0da7}

\bibitem[{{Huyan} {et~al.}(2023){Huyan}, {Kulkarni}, {Poudel}, {Tejos},
  {P{\'e}roux}, \& {Lopez}}]{2023ApJ...954L..19H}
{Huyan}, J., {Kulkarni}, V.~P., {Poudel}, S., {et~al.} 2023, \apjl, 954, L19,
  \dodoi{10.3847/2041-8213/aceefe}

\bibitem[{{Jenkins}(2009)}]{2009ApJ...700.1299J}
{Jenkins}, E.~B. 2009, \apj, 700, 1299, \dodoi{10.1088/0004-637X/700/2/1299}

\bibitem[{{Kelson}(2003)}]{2003PASP..115..688K}
{Kelson}, D.~D. 2003, \pasp, 115, 688, \dodoi{10.1086/375502}

\bibitem[{{Kelson} {et~al.}(2000){Kelson}, {Illingworth}, {van Dokkum}, \&
  {Franx}}]{2000ApJ...531..159K}
{Kelson}, D.~D., {Illingworth}, G.~D., {van Dokkum}, P.~G., \& {Franx}, M.
  2000, \apj, 531, 159, \dodoi{10.1086/308445}

\bibitem[{{Khaire} \& {Srianand}(2019)}]{2019MNRAS.484.4174K}
{Khaire}, V., \& {Srianand}, R. 2019, \mnras, 484, 4174,
  \dodoi{10.1093/mnras/stz174}

\bibitem[{{Konstantopoulou} {et~al.}(2024){Konstantopoulou}, {De Cia},
  {Ledoux}, {Krogager}, {Mattsson}, {Watson}, {Heintz}, {P{\'e}roux},
  {Noterdaeme}, {Andersen}, {Fynbo}, {Jermann}, \&
  {Ramburuth-Hurt}}]{2024A&A...681A..64K}
{Konstantopoulou}, C., {De Cia}, A., {Ledoux}, C., {et~al.} 2024, \aap, 681,
  A64, \dodoi{10.1051/0004-6361/202347171}

\bibitem[{{Krogager}(2018)}]{2018ascl.soft11016K}
{Krogager}, J.-K. 2018, {VoigtFit: Absorption line fitting for Voigt profiles},
  Astrophysics Source Code Library, record ascl:1811.016

\bibitem[{{Kulkarni} {et~al.}(2014){Kulkarni}, {Hennawi}, {Rollinde}, \&
  {Vangioni}}]{2014ApJ...787...64K}
{Kulkarni}, G., {Hennawi}, J.~F., {Rollinde}, E., \& {Vangioni}, E. 2014, \apj,
  787, 64, \dodoi{10.1088/0004-637X/787/1/64}

\bibitem[{{Kulkarni} {et~al.}(2013){Kulkarni}, {Rollinde}, {Hennawi}, \&
  {Vangioni}}]{2013ApJ...772...93K}
{Kulkarni}, G., {Rollinde}, E., {Hennawi}, J.~F., \& {Vangioni}, E. 2013, \apj,
  772, 93, \dodoi{10.1088/0004-637X/772/2/93}

\bibitem[{{Kulkarni} \& {Fall}(2002)}]{2002ApJ...580..732K}
{Kulkarni}, V.~P., \& {Fall}, S.~M. 2002, \apj, 580, 732,
  \dodoi{10.1086/343855}

\bibitem[{{Kulkarni} {et~al.}(2005){Kulkarni}, {Fall}, {Lauroesch}, {York},
  {Welty}, {Khare}, \& {Truran}}]{2005ApJ...618...68K}
{Kulkarni}, V.~P., {Fall}, S.~M., {Lauroesch}, J.~T., {et~al.} 2005, \apj, 618,
  68, \dodoi{10.1086/425956}

\bibitem[{{L{\'o}pez} {et~al.}(2016){L{\'o}pez}, {D'Odorico}, {Ellison},
  {Becker}, {Christensen}, {Cupani}, {Denney}, {P{\^a}ris}, {Worseck}, {Berg},
  {Cristiani}, {Dessauges-Zavadsky}, {Haehnelt}, {Hamann}, {Hennawi},
  {Ir{\v{s}}i{\v{c}}}, {Kim}, {L{\'o}pez}, {Lund Saust}, {M{\'e}nard},
  {Perrotta}, {Prochaska}, {S{\'a}nchez-Ram{\'\i}rez}, {Vestergaard}, {Viel},
  \& {Wisotzki}}]{2016A&A...594A..91L}
{L{\'o}pez}, S., {D'Odorico}, V., {Ellison}, S.~L., {et~al.} 2016, \aap, 594,
  A91, \dodoi{10.1051/0004-6361/201628161}

\bibitem[{{Maio} \& {Tescari}(2015)}]{2015MNRAS.453.3798M}
{Maio}, U., \& {Tescari}, E. 2015, \mnras, 453, 3798,
  \dodoi{10.1093/mnras/stv1714}

\bibitem[{{Meiring} {et~al.}(2006){Meiring}, {Kulkarni}, {Khare}, {Bechtold},
  {York}, {Cui}, {Lauroesch}, {Crotts}, \& {Nakamura}}]{2006MNRAS.370...43M}
{Meiring}, J.~D., {Kulkarni}, V.~P., {Khare}, P., {et~al.} 2006, \mnras, 370,
  43, \dodoi{10.1111/j.1365-2966.2006.10500.x}

\bibitem[{{Morrison} {et~al.}(2016){Morrison}, {Kulkarni}, {Som}, {DeMarcy},
  {Quiret}, \& {P{\'e}roux}}]{2016ApJ...830..158M}
{Morrison}, S., {Kulkarni}, V.~P., {Som}, D., {et~al.} 2016, \apj, 830, 158,
  \dodoi{10.3847/0004-637X/830/2/158}

\bibitem[{{P{\'e}roux} \& {Howk}(2020)}]{2020ARA&A..58..363P}
{P{\'e}roux}, C., \& {Howk}, J.~C. 2020, \araa, 58, 363,
  \dodoi{10.1146/annurev-astro-021820-120014}

\bibitem[{{P{\'e}roux} {et~al.}(2003){P{\'e}roux}, {McMahon},
  {Storrie-Lombardi}, \& {Irwin}}]{2003MNRAS.346.1103P}
{P{\'e}roux}, C., {McMahon}, R.~G., {Storrie-Lombardi}, L.~J., \& {Irwin},
  M.~J. 2003, \mnras, 346, 1103, \dodoi{10.1111/j.1365-2966.2003.07129.x}

\bibitem[{{Poudel} {et~al.}(2020){Poudel}, {Kulkarni}, {Cashman}, {Frye},
  {P{\'e}roux}, {Rahmani}, \& {Quiret}}]{2020MNRAS.491.1008P}
{Poudel}, S., {Kulkarni}, V.~P., {Cashman}, F.~H., {et~al.} 2020, \mnras, 491,
  1008, \dodoi{10.1093/mnras/stz3000}

\bibitem[{{Poudel} {et~al.}(2018){Poudel}, {Kulkarni}, {Morrison},
  {P{\'e}roux}, {Som}, {Rahmani}, \& {Quiret}}]{2018MNRAS.473.3559P}
{Poudel}, S., {Kulkarni}, V.~P., {Morrison}, S., {et~al.} 2018, \mnras, 473,
  3559, \dodoi{10.1093/mnras/stx2607}

\bibitem[{{Prochaska} {et~al.}(2007){Prochaska}, {Chen}, {Dessauges-Zavadsky},
  \& {Bloom}}]{2007ApJ...666..267P}
{Prochaska}, J.~X., {Chen}, H.-W., {Dessauges-Zavadsky}, M., \& {Bloom}, J.~S.
  2007, \apj, 666, 267, \dodoi{10.1086/520042}

\bibitem[{{Prochaska} {et~al.}(2003){Prochaska}, {Gawiser}, {Wolfe}, {Castro},
  \& {Djorgovski}}]{2003ApJ...595L...9P}
{Prochaska}, J.~X., {Gawiser}, E., {Wolfe}, A.~M., {Castro}, S., \&
  {Djorgovski}, S.~G. 2003, \apjl, 595, L9, \dodoi{10.1086/378945}

\bibitem[{{Rafelski} {et~al.}(2014){Rafelski}, {Neeleman}, {Fumagalli},
  {Wolfe}, \& {Prochaska}}]{2014ApJ...782L..29R}
{Rafelski}, M., {Neeleman}, M., {Fumagalli}, M., {Wolfe}, A.~M., \&
  {Prochaska}, J.~X. 2014, \apjl, 782, L29, \dodoi{10.1088/2041-8205/782/2/L29}

\bibitem[{{Rafelski} {et~al.}(2012){Rafelski}, {Wolfe}, {Prochaska},
  {Neeleman}, \& {Mendez}}]{2012ApJ...755...89R}
{Rafelski}, M., {Wolfe}, A.~M., {Prochaska}, J.~X., {Neeleman}, M., \&
  {Mendez}, A.~J. 2012, \apj, 755, 89, \dodoi{10.1088/0004-637X/755/2/89}

\bibitem[{{Robertson} {et~al.}(2024){Robertson}, {Johnson}, {Tacchella},
  {Eisenstein}, {Hainline}, {Arribas}, {Baker}, {Bunker}, {Carniani},
  {Cargile}, {Carreira}, {Charlot}, {Chevallard}, {Curti}, {Curtis-Lake},
  {D'Eugenio}, {Egami}, {Hausen}, {Helton}, {Jakobsen}, {Ji}, {Jones},
  {Maiolino}, {Maseda}, {Nelson}, {P{\'e}rez-Gonz{\'a}lez}, {Pusk{\'a}s},
  {Rieke}, {Smit}, {Sun}, {{\"U}bler}, {Whitler}, {Williams}, {Willmer},
  {Willott}, \& {Witstok}}]{2024ApJ...970...31R}
{Robertson}, B., {Johnson}, B.~D., {Tacchella}, S., {et~al.} 2024, \apj, 970,
  31, \dodoi{10.3847/1538-4357/ad463d}

\bibitem[{{Sebastian} {et~al.}(2024){Sebastian}, {Ryan-Weber}, {Davies},
  {Becker}, {Keating}, {D'Odorico}, {Meyer}, {Bosman}, {Cupani}, {Kulkarni},
  {Haehnelt}, {Lai}, {Eilers}, {Bischetti}, \&
  {Gallerani}}]{2024MNRAS.530.1829S}
{Sebastian}, A.~M., {Ryan-Weber}, E., {Davies}, R.~L., {et~al.} 2024, \mnras,
  530, 1829, \dodoi{10.1093/mnras/stae789}

\bibitem[{{Sodini} {et~al.}(2024){Sodini}, {D'Odorico}, {Salvadori}, {Vanni},
  {Bischetti}, {Cupani}, {Davies}, {Becker}, {Ba{\~n}ados}, {Bosman}, {Davies},
  {Paolo Farina}, {Ferrara}, {Keating}, {Kulkarni}, {Lai}, {Ryan-Weber}, {Maria
  Sebastian}, \& {Walter}}]{2024A&A...687A.314S}
{Sodini}, A., {D'Odorico}, V., {Salvadori}, S., {et~al.} 2024, \aap, 687, A314,
  \dodoi{10.1051/0004-6361/202349062}

\bibitem[{{Som} {et~al.}(2015){Som}, {Kulkarni}, {Meiring}, {York},
  {P{\'e}roux}, {Lauroesch}, {Aller}, \& {Khare}}]{2015ApJ...806...25S}
{Som}, D., {Kulkarni}, V.~P., {Meiring}, J., {et~al.} 2015, \apj, 806, 25,
  \dodoi{10.1088/0004-637X/806/1/25}

\bibitem[{{Tacchella} {et~al.}(2023){Tacchella}, {Eisenstein}, {Hainline},
  {Johnson}, {Baker}, {Helton}, {Robertson}, {Suess}, {Chen}, {Nelson},
  {Pusk{\'a}s}, {Sun}, {Alberts}, {Egami}, {Hausen}, {Rieke}, {Rieke},
  {Shivaei}, {Williams}, {Willmer}, {Bunker}, {Cameron}, {Carniani}, {Charlot},
  {Curti}, {Curtis-Lake}, {Looser}, {Maiolino}, {Maseda}, {Rawle}, {Rix},
  {Smit}, {{\"U}bler}, {Willott}, {Witstok}, {Baum}, {Bhatawdekar}, {Boyett},
  {Danhaive}, {de Graaff}, {Endsley}, {Ji}, {Lyu}, {Sandles}, {Saxena},
  {Scholtz}, {Topping}, \& {Whitler}}]{2023ApJ...952...74T}
{Tacchella}, S., {Eisenstein}, D.~J., {Hainline}, K., {et~al.} 2023, \apj, 952,
  74, \dodoi{10.3847/1538-4357/acdbc6}

\bibitem[{{Tripodi} {et~al.}(2024){Tripodi}, {D'Eugenio}, {Maiolino}, {Curti},
  {Scholtz}, {Tacchella}, {Marconcini}, {Bunker}, {Trussler}, {Cameron},
  {Arribas}, {Baker}, {Brada{\v{c}}}, {Carniani}, {Charlot}, {Ji}, {Ji},
  {Robertson}, {{\"U}bler}, {Venturi}, {Willmer}, \&
  {Witstok}}]{2024A&A...692A.184T}
{Tripodi}, R., {D'Eugenio}, F., {Maiolino}, R., {et~al.} 2024, \aap, 692, A184,
  \dodoi{10.1051/0004-6361/202449980}

\bibitem[{{Vanni} {et~al.}(2023){Vanni}, {Salvadori}, {Sk{\'u}lad{\'o}ttir},
  {Rossi}, \& {Koutsouridou}}]{2023MNRAS.526.2620V}
{Vanni}, I., {Salvadori}, S., {Sk{\'u}lad{\'o}ttir}, {\'A}., {Rossi}, M., \&
  {Koutsouridou}, I. 2023, \mnras, 526, 2620, \dodoi{10.1093/mnras/stad2910}

\bibitem[{{Wang} {et~al.}(2016){Wang}, {Wu}, {Fan}, {Yang}, {Yi}, {Bian},
  {McGreer}, {Yang}, {Ai}, {Dong}, {Zuo}, {Jiang}, {Green}, {Wang}, {Cai},
  {Wang}, \& {Yue}}]{2016ApJ...819...24W}
{Wang}, F., {Wu}, X.-B., {Fan}, X., {et~al.} 2016, \apj, 819, 24,
  \dodoi{10.3847/0004-637X/819/1/24}

\bibitem[{{Welsh} {et~al.}(2019){Welsh}, {Cooke}, \&
  {Fumagalli}}]{2019MNRAS.487.3363W}
{Welsh}, L., {Cooke}, R., \& {Fumagalli}, M. 2019, \mnras, 487, 3363,
  \dodoi{10.1093/mnras/stz1526}

\bibitem[{{Welsh} {et~al.}(2022){Welsh}, {Cooke}, {Fumagalli}, \&
  {Pettini}}]{2022ApJ...929..158W}
{Welsh}, L., {Cooke}, R., {Fumagalli}, M., \& {Pettini}, M. 2022, \apj, 929,
  158, \dodoi{10.3847/1538-4357/ac4503}

\bibitem[{{Wolf} {et~al.}(2020){Wolf}, {Hon}, {Bian}, {Onken}, {Alonzi},
  {Bessell}, {Li}, {Schmidt}, \& {Tisserand}}]{2020MNRAS.491.1970W}
{Wolf}, C., {Hon}, W.~J., {Bian}, F., {et~al.} 2020, \mnras, 491, 1970,
  \dodoi{10.1093/mnras/stz2955}

\bibitem[{{Yang} {et~al.}(2016){Yang}, {Wang}, {Wu}, {Fan}, {McGreer}, {Bian},
  {Yi}, {Yang}, {Ai}, {Dong}, {Zuo}, {Green}, {Jiang}, {Wang}, {Wang}, \&
  {Yue}}]{2016ApJ...829...33Y}
{Yang}, J., {Wang}, F., {Wu}, X.-B., {et~al.} 2016, \apj, 829, 33,
  \dodoi{10.3847/0004-637X/829/1/33}

\bibitem[{{Yang} {et~al.}(2019){Yang}, {Wang}, {Fan}, {Wu}, {Bian},
  {Ba{\~n}ados}, {Yue}, {Schindler}, {Yang}, {Jiang}, {McGreer}, {Green}, \&
  {Dye}}]{2019ApJ...871..199Y}
{Yang}, J., {Wang}, F., {Fan}, X., {et~al.} 2019, \apj, 871, 199,
  \dodoi{10.3847/1538-4357/aaf858}

\bibitem[{{Yates} {et~al.}(2021){Yates}, {P{\'e}roux}, \&
  {Nelson}}]{2021MNRAS.508.3535Y}
{Yates}, R.~M., {P{\'e}roux}, C., \& {Nelson}, D. 2021, \mnras, 508, 3535,
  \dodoi{10.1093/mnras/stab2837}

\end{thebibliography}

\begin{center}
\begin{longtable}{cccccc}
\caption{Results of the Voigt profile fitting for the DLAs} \label{tab:absorption-catlog} \\

\hline \multicolumn{1}{c}{\textbf{QSO Name}} & \multicolumn{1}{c}{\textbf{ion}} & \multicolumn{1}{c}{\textbf{$z_{abs}$}} & \multicolumn{1}{c}{\textbf{b (km/s)}} & \multicolumn{1}{c}{\textbf{logN$_{HI}$/cm$^{-2}$}}  & \multicolumn{1}{c}{\textbf{label}}  \\ \hline 
\endfirsthead

\multicolumn{6}{c}%
{{\bfseries \tablename\ \thetable{} -- continued from previous page}} \\
\hline \multicolumn{1}{c}{\textbf{QSO Name}} & \multicolumn{1}{c}{\textbf{ion}} & \multicolumn{1}{c}{\textbf{$z_{abs}$}} & \multicolumn{1}{c}{\textbf{b (km/s)}} & \multicolumn{1}{c}{\textbf{logN$_{HI}$/cm$^{-2}$}}   & \multicolumn{1}{c}{\textbf{label}} \\ \hline 
\endhead

\hline \multicolumn{6}{r}{{Continued on next page}} \\ \hline
\endfoot

\hline \hline
\endlastfoot

J0007-5701 & HI & 4.24168 & 4.19 & $21.30_{-0.10}^{+0.10}$ & component-b  \\
& SII & 4.240630 & $5.10_{-0.27}^{+0.28}$ & $13.51_{-0.13}^{+0.10}$ & component-a  \\
&    & 4.241680 & $4.19_{-0.14}^{+0.16}$ & $14.15_{-0.03}^{+0.03}$ & component-b  \\
&    & 4.242220 & $4.49_{-0.16}^{+0.18}$ & $14.70_{-0.03}^{+0.03}$ & component-c  \\
&    & 4.242465 & $6.57_{-0.34}^{+0.36}$ & $12.25_{-1.49}^{+0.83}$ & component-d  \\
& CIIa & 4.241680 & $5.10_{-0.27}^{+0.28}$ & $13.42_{-0.03}^{+0.03}$ & component-b  \\
&    & 4.242220 & $4.19_{-0.14}^{+0.16}$ & $13.30_{-0.04}^{+0.04}$ & component-c  \\
& FeII & 4.240630 & $5.10_{-0.27}^{+0.28}$ & $13.44_{-0.04}^{+0.03}$ & component-a  \\
&    & 4.241680 & $4.19_{-0.14}^{+0.16}$ & $14.20_{-0.05}^{+0.06}$ & component-b  \\
&    & 4.242220 & $4.49_{-0.16}^{+0.18}$ & $15.06_{-0.12}^{+0.13}$ & component-c  \\
&    & 4.242465 & $6.57_{-0.34}^{+0.36}$ & $13.40_{-0.04}^{+0.04}$ & component-d  \\
& SiII & 4.240630 & $5.10_{-0.27}^{+0.28}$ & $13.86_{-0.04}^{+0.04}$ & component-a  \\
&    & 4.241680 & $4.19_{-0.14}^{+0.16}$ & $\ge 14.84$ & component-b  \\
&    & 4.242220 & $4.49_{-0.16}^{+0.18}$ & $\ge 15.04$ & component-c  \\
&    & 4.242465 & $6.57_{-0.34}^{+0.36}$ & $13.73_{-0.02}^{+0.02}$ & component-d  \\
& CII & 4.240630 & $5.10$ & $\ge 16.04$ & component-a  \\
&    & 4.241680 & $4.19$ & $\ge 16.33$ & component-b  \\
&    & 4.242220 & $4.49$ & $\ge 16.31$ & component-c  \\
&    & 4.242465 & $6.57$ & $\ge 14.55$ & component-d  \\
& OI & 4.240630 & $5.10$ & $\ge 16.34$ & component-a  \\
&    & 4.241680 & $4.19$ & $\ge 16.93$ & component-b  \\
&    & 4.242220 & $4.49$ & $\ge 16.51$ & component-c  \\
&    & 4.242465 & $6.57$ & $\ge 14.65$ & component-d  \\
& OI & 4.240630 & $5.10$ & $\ge 16.34$ & component-a  \\
&    & 4.241680 & $4.19$ & $\ge 16.93$ & component-b  \\
&    & 4.242220 & $4.49$ & $\ge 16.51$ & component-c  \\
&    & 4.242465 & $6.57$ & $\ge 14.65$ & component-d  \\
& CIV & 4.241138 & $18.59_{-0.54}^{+0.54}$ & $13.69_{-0.02}^{+0.01}$ & component-a  \\
&    & 4.241330 & $10.99_{-0.79}^{+0.85}$ & $13.33_{-0.03}^{+0.03}$ & component-b  \\
&    & 4.242152 & $20.08_{-0.47}^{+0.46}$ & $13.89_{-0.01}^{+0.01}$ & component-c  \\
&    & 4.242554 & $14.65_{-0.82}^{+0.83}$ & $13.20_{-0.05}^{+0.04}$ & component-d  \\
& SiIV & 4.240946 & $14.65_{-0.82}^{+0.83}$ & $12.99_{-0.02}^{+0.02}$ & component-a$^{\prime}$  \\
&    & 4.241330 & $18.59_{-0.54}^{+0.54}$ & $12.66_{-0.03}^{+0.03}$ & component-b  \\
&    & 4.242152 & $10.99_{-0.79}^{+0.85}$ & $13.07_{-0.01}^{+0.01}$ & component-c  \\
&    & 4.242554 & $20.08_{-0.47}^{+0.46}$ & $\le 11.70$ & component-d  \\

\hline
J0838-0440 & HI &  4.713615 & 6.36 & $20.80_{-0.10}^{+0.10}$ & component-f  \\
& SII & 4.712070 & $12.42_{-1.72}^{+1.81}$ & $14.30_{-0.08}^{+0.06}$ & component-c  \\
&    & 4.712685 & $3.27_{-1.11}^{+1.96}$ & $13.98_{-0.27}^{+0.17}$ & component-d  \\
&    & 4.713112 & $25.30_{-5.80}^{+5.37}$ & $14.57_{-0.11}^{+0.11}$ & component-e  \\
&    & 4.713615 & $6.36_{-0.66}^{+0.67}$ & $15.09_{-0.04}^{+0.05}$ & component-f  \\
& FeII & 4.712070 & $12.42_{-1.72}^{+1.81}$ & $14.13_{-0.04}^{+0.04}$ & component-c  \\
&    & 4.712685 & $3.27_{-1.11}^{+1.96}$ & $14.94_{-0.77}^{+1.32}$ & component-d  \\
&    & 4.713112 & $25.30_{-5.80}^{+5.37}$ & $14.20_{-0.10}^{+0.08}$ & component-e  \\
&    & 4.713615 & $6.36_{-0.66}^{+0.67}$ & $14.88_{-0.23}^{+0.30}$ & component-f  \\
& CII & 4.711427 & $10.00$ & $\ge 13.95$ & component-a  \\
&    & 4.711808 & $10.00$ & $\ge 14.16$ & component-b  \\
&    & 4.712070 & $12.42$ & $\ge 14.50$ & component-c  \\
&    & 4.712685 & $3.27$ & $\ge 14.58$ & component-d  \\
&    & 4.713112 & $25.31$ & $\ge 14.82$ & component-e  \\
&    & 4.713615 & $6.36$ & $\ge 13.89$ & component-f  \\
&    & 4.713848 & $10.00$ & $\ge 14.30$ & component-g  \\
&    & 4.714134 & $10.00$ & $\ge 14.20$ & component-h  \\
& OI & 4.711427 & $10.00$ & $\ge 13.30$ & component-a  \\
&    & 4.711808 & $10.00$ & $\ge 14.70$ & component-b  \\
&    & 4.712070 & $12.42$ & $\ge 15.20$ & component-c  \\
&    & 4.712685 & $3.27$ & $\ge 15.58$ & component-d  \\
&    & 4.713112 & $25.31$ & $\ge 15.06$ & component-e  \\
&    & 4.713615 & $6.36$ & $\ge 15.19$ & component-f  \\
&    & 4.713848 & $10.00$ & $\ge 14.30$ & component-g  \\
&    & 4.714134 & $10.00$ & $\ge 14.20$ & component-h  \\
& SiII & 4.711427 & $10.00$ & $\ge 13.00$ & component-a  \\
&    & 4.711808 & $10.00$ & $\ge 14.00$ & component-b  \\
&    & 4.712070 & $12.42$ & $\ge 14.43$ & component-c  \\
&    & 4.712685 & $3.27$ & $\ge 14.74$ & component-d  \\
&    & 4.713112 & $25.31$ & $\ge 14.82$ & component-e  \\
&    & 4.713615 & $6.36$ & $\ge 14.58$ & component-f  \\
&    & 4.713848 & $10.00$ & $\ge 13.80$ & component-g  \\
&    & 4.714134 & $10.00$ & $\ge 13.50$ & component-h  \\

\hline
J1129-0142 & HI & 4.494940 & 10.0 & $20.80_{-0.10}^{+0.10}$  & component-a  \\
& OI & 4.494757 & $21.62_{-0.39}^{+0.39}$ & $15.31_{-0.03}^{+0.03}$ & component-a  \\
&    & 4.495453 & $2.65_{-0.17}^{+0.17}$ & $16.44_{-0.25}^{+0.21}$ & component-b  \\
&    & 4.496131 & $21.48_{-1.04}^{+1.00}$ & $14.24_{-0.02}^{+0.02}$ & component-c  \\
&    & 4.497048 & $4.39_{-0.67}^{+0.97}$ & $13.19_{-0.12}^{+0.10}$ & component-d  \\
&    & 4.497598 & $37.12_{-1.99}^{+2.02}$ & $13.97_{-0.03}^{+0.03}$ & component-e  \\
&    & 4.497836 & $0.26_{-0.06}^{+0.07}$ & $12.29_{-1.55}^{+1.47}$ & component-f  \\
& SiII & 4.494757 & $21.62_{-0.39}^{+0.39}$ & $14.32_{-0.01}^{+0.01}$ & component-a  \\
&    & 4.495453 & $2.65_{-0.17}^{+0.17}$ & $14.44_{-0.15}^{+0.20}$ & component-b  \\
&    & 4.496131 & $21.48_{-1.04}^{+1.00}$ & $13.42_{-0.03}^{+0.03}$ & component-c  \\
&    & 4.497048 & $4.39_{-0.67}^{+0.97}$ & $12.54_{-0.23}^{+0.15}$ & component-d  \\
&    & 4.497598 & $37.12_{-1.99}^{+2.02}$ & $13.78_{-0.02}^{+0.02}$ & component-e  \\
&    & 4.497836 & $0.26_{-0.06}^{+0.07}$ & $13.11_{-0.31}^{+0.32}$ & component-f  \\
& CII & 4.494757 & $21.62_{-0.39}^{+0.39}$ & $15.01_{-0.05}^{+0.05}$ & component-a  \\
&    & 4.495453 & $2.65_{-0.17}^{+0.17}$ & $17.86_{-0.30}^{+0.31}$ & component-b  \\
&    & 4.496131 & $21.48_{-1.04}^{+1.00}$ & $14.58_{-0.05}^{+0.06}$ & component-c  \\
&    & 4.497048 & $4.39_{-0.67}^{+0.97}$ & $15.22_{-0.60}^{+0.68}$ & component-d  \\
&    & 4.497598 & $37.12_{-1.99}^{+2.02}$ & $11.30_{-0.89}^{+0.99}$ & component-e  \\
&    & 4.497836 & $0.26_{-0.06}^{+0.07}$ & $18.00_{-0.02}^{+0.02}$ & component-f  \\
& CIV & 4.492429 & $3.64_{-0.48}^{+0.95}$ & $13.37_{-0.08}^{+0.08}$ & component-a  \\
&    & 4.493125 & $12.72_{-6.59}^{+9.86}$ & $12.96_{-0.13}^{+0.11}$ & component-b  \\
&    & 4.493969 & $19.20_{-4.13}^{+6.42}$ & $13.33_{-0.07}^{+0.08}$ & component-c  \\
&    & 4.494848 & $14.81_{-0.63}^{+0.62}$ & $14.19_{-0.03}^{+0.03}$ & component-d  \\
&    & 4.496131 & $24.85_{-4.57}^{+5.05}$ & $13.24_{-0.06}^{+0.06}$ & component-e  \\
&    & 4.498111 & $29.59_{-1.66}^{+1.70}$ & $13.67_{-0.03}^{+0.03}$ & component-f  \\
&    & 4.498771 & $34.71_{-4.46}^{+3.98}$ & $13.39_{-0.08}^{+0.06}$ & component-g  \\
&    & 4.499247 & $20.46_{-13.90}^{+19.73}$ & $11.38_{-0.95}^{+0.92}$ & component-h  \\
& SiIV & 4.492429 & $3.64_{-0.48}^{+0.95}$ & $12.75_{-0.11}^{+0.10}$ & component-a  \\
&    & 4.493125 & $12.72_{-6.59}^{+9.86}$ & $11.34_{-0.91}^{+0.73}$ & component-b  \\
&    & 4.493969 & $19.20_{-4.13}^{+6.42}$ & $12.69_{-0.16}^{+0.13}$ & component-c  \\
&    & 4.494848 & $14.81_{-0.63}^{+0.62}$ & $13.80_{-0.03}^{+0.03}$ & component-d  \\
&    & 4.496131 & $24.85_{-4.57}^{+5.05}$ & $12.45_{-0.67}^{+0.21}$ & component-e  \\
&    & 4.498111 & $29.59_{-1.66}^{+1.70}$ & $13.75_{-0.04}^{+0.04}$ & component-f  \\
&    & 4.498771 & $34.71_{-4.46}^{+3.98}$ & $11.26_{-0.86}^{+0.89}$ & component-g  \\
&    & 4.499247 & $20.46_{-13.90}^{+19.73}$ & $11.34_{-0.91}^{+0.77}$ & component-h  \\

\hline
J1135-1526 & HI & 4.555380 & 13.55 & $20.30_{-0.10}^{+0.10}$ & component-d  \\
& CII & 4.552878 & $16.16_{-1.73}^{+1.78}$ & $13.80_{-0.07}^{+0.05}$ & component-a  \\
&    & 4.553805 & $30.74_{-3.19}^{+4.22}$ & $14.33_{-0.03}^{+0.03}$ & component-b  \\
&    & 4.554731 & $11.03_{-2.16}^{+1.87}$ & $14.27_{-0.10}^{+0.21}$ & component-c  \\
&    & 4.555380 & $13.55_{-1.17}^{+1.25}$ & $14.42_{-0.16}^{+0.19}$ & component-d  \\
&    & 4.555806 & $19.46_{-1.19}^{+1.21}$ & $14.79_{-0.10}^{+0.13}$ & component-e  \\
& SiII & 4.552878 & $16.16_{-1.73}^{+1.78}$ & $13.13_{-0.12}^{+0.09}$ & component-a  \\
&    & 4.553805 & $30.74_{-3.19}^{+4.22}$ & $13.61_{-0.05}^{+0.05}$ & component-b  \\
&    & 4.554731 & $11.03_{-2.16}^{+1.87}$ & $13.43_{-0.07}^{+0.06}$ & component-c  \\
&    & 4.555380 & $13.55_{-1.17}^{+1.25}$ & $13.91_{-0.07}^{+0.07}$ & component-d  \\
&    & 4.555806 & $19.46_{-1.19}^{+1.21}$ & $14.08_{-0.04}^{+0.04}$ & component-e  \\
& SII & 4.555380 & $13.55_{-1.17}^{+1.25}$ & $14.38_{-0.05}^{+0.04}$ & component-d  \\
& FeII & 4.555380 & $13.55_{-1.17}^{+1.25}$ & $13.93_{-0.04}^{+0.04}$ & component-d  \\
& OI & 4.552878 & $19.46$ & $14.20_{-0.05}^{+0.05}$ & component-a  \\
&    & 4.553805 & $13.55$ & $14.88_{-0.03}^{+0.03}$ & component-b  \\
&    & 4.554731 & $11.03$ & $14.34_{-0.06}^{+0.07}$ & component-c  \\
&    & 4.555380 & $30.74$ & $15.23_{-0.13}^{+0.15}$ & component-d  \\
&    & 4.555806 & $16.16$ & $15.21_{-0.05}^{+0.06}$ & component-e  \\

\hline
J1135-1526 & HI & 4.346010 & 2.90 & $19.90_{-0.10}^{+0.10}$ & component-b  \\
& CII & 4.345796 & $11.52_{-2.76}^{+3.01}$ & $13.15_{-0.10}^{+0.09}$ & component-a  \\
&    & 4.346010 & $2.90_{-0.48}^{+0.82}$ & $13.81_{-0.22}^{+0.35}$ & component-b  \\
&    & 4.346277 & $10.03_{-1.90}^{+1.95}$ & $13.37_{-0.06}^{+0.06}$ & component-c  \\
& OI & 4.346010 & $2.90_{-0.48}^{+0.82}$ & $15.98_{-0.84}^{+0.50}$ & component-b  \\
& SiII & 4.345796 & $11.52_{-2.76}^{+3.01}$ & $12.65_{-0.73}^{+0.22}$ & component-a  \\
&    & 4.346010 & $2.90_{-0.48}^{+0.82}$ & $12.04_{-1.28}^{+0.56}$ & component-b  \\

\hline
J1336–1830 & HI & 4.518762 & 4.93 & $21.00_{-0.10}^{+0.10}$ & component-a  \\
& CII & 4.518762 & $4.93_{-0.24}^{+0.26}$ & $16.56_{-0.20}^{+0.16}$ & component-a  \\
& OI & 4.518762 & $4.93_{-0.24}^{+0.26}$ & $16.87_{-0.19}^{+0.16}$ & component-a  \\
& SiII & 4.518762 & $4.93_{-0.24}^{+0.26}$ & $14.34_{-0.08}^{+0.09}$ & component-a  \\
& FeII & 4.518762 & $4.93_{-0.24}^{+0.26}$ & $13.68_{-0.04}^{+0.04}$ & component-a   \\

\hline
J1336–1830 & HI & 4.286280 & 5.00 & $20.60_{-0.10}^{+0.10}$ & component-b  \\
& SiII & 4.285821 & $5.35_{-0.52}^{+0.70}$ & $13.95_{-0.11}^{+0.14}$ & component-a  \\
&    & 4.286315 & $8.40_{-1.31}^{+1.11}$ & $13.84_{-0.08}^{+0.09}$ & component-b  \\
&    & 4.286721 & $16.26_{-6.68}^{+6.67}$ & $13.62_{-0.12}^{+0.16}$ & component-c  \\
&    & 4.287056 & $12.49_{-1.46}^{+1.00}$ & $12.97_{-1.96}^{+0.23}$ & component-d  \\
& CII & 4.285821 & $5.35_{-0.52}^{+0.70}$ & $15.77_{-0.48}^{+0.43}$ & component-a  \\
&    & 4.286315 & $8.40_{-1.31}^{+1.11}$ & $14.47_{-0.13}^{+0.25}$ & component-b  \\
&    & 4.286721 & $16.26_{-6.68}^{+6.67}$ & $14.31_{-0.14}^{+0.11}$ & component-c  \\
&    & 4.287056 & $12.49_{-1.46}^{+1.00}$ & $13.92_{-0.30}^{+0.10}$ & component-d  \\
& FeII & 4.285821 & $5.35_{-0.52}^{+0.70}$ & $12.92_{-0.22}^{+0.14}$ & component-a  \\
&    & 4.286315 & $8.40_{-1.31}^{+1.11}$ & $13.40_{-0.09}^{+0.08}$ & component-b  \\
&    & 4.286721 & $16.26_{-6.68}^{+6.67}$ & $13.28_{-0.14}^{+0.13}$ & component-c  \\
&    & 4.287056 & $12.49_{-1.46}^{+1.00}$ & $11.34_{-0.91}^{+0.96}$ & component-d  \\
& OI & 4.285821 & $5.35$ & $\ge 15.58$ & component-a  \\
&    & 4.286315 & $8.40$ & $\ge 15.16$ & component-b  \\
&    & 4.286721 & $16.26$ & $\ge 14.80$ & component-c  \\
&    & 4.287056 & $12.49$ & $\ge 14.32$ & component-d  \\

\hline
JJ2202+1509 & HI & 4.948650 & 8.97 & $20.20_{-0.10}^{+0.10}$ & component-b  \\
& CII & 4.948133 & $7.27_{-0.44}^{+0.45}$ & $14.28_{-0.14}^{+0.19}$ & component-a  \\
&    & 4.948658 & $8.97_{-0.70}^{+0.80}$ & $14.34_{-0.13}^{+0.18}$ & component-b  \\
&    & 4.949153 & $4.87_{-1.60}^{+1.83}$ & $13.28_{-0.13}^{+0.13}$ & component-c  \\
& OI & 4.948133 & $7.27_{-0.44}^{+0.45}$ & $14.46_{-0.10}^{+0.13}$ & component-a  \\
&    & 4.948658 & $8.97_{-0.70}^{+0.80}$ & $14.37_{-0.07}^{+0.08}$ & component-b  \\
& SiII & 4.948133 & $7.27_{-0.44}^{+0.45}$ & $13.56_{-0.06}^{+0.06}$ & component-a  \\
&    & 4.948658 & $8.97_{-0.70}^{+0.80}$ & $13.26_{-0.05}^{+0.06}$ & component-b  \\
&    & 4.949153 & $4.87_{-1.60}^{+1.83}$ & $12.43_{-0.07}^{+0.11}$ & component-c  \\

\hline
J2328+0217 & HI & 4.889080  & 3.30 & $20.20_{-0.10}^{+0.10}$ & component-a  \\
& OI & 4.889135 & $3.30_{-0.22}^{+0.48}$ & $15.38_{-0.43}^{+0.36}$ & component-a  \\
& SiII & 4.889135 & $3.30_{-0.22}^{+0.48}$ & $13.12_{-0.23}^{+0.25}$ & component-a  \\
& CII & 4.889135 & $3.30_{-0.22}^{+0.48}$ & $13.74_{-0.14}^{+0.16}$ & component-a  \\
& CIV & 4.887148 & $30.72_{-5.21}^{+6.64}$ & $13.68_{-0.08}^{+0.07}$ & component-a  \\
&    & 4.889347 & $21.10_{-2.22}^{+2.43}$ & $13.77_{-0.07}^{+0.06}$ & component-b  \\
& SiIV & 4.887148 & $30.72_{-5.21}^{+6.64}$ & $12.40_{-0.18}^{+0.13}$ & component-a  \\
&    & 4.889347 & $21.10_{-2.22}^{+2.43}$ & $12.97_{-0.04}^{+0.04}$ & component-b  \\

\hline
J2328+0217 & HI & 4.655690  & 17.72 & $20.30_{-0.10}^{+0.10}$ & component-b  \\
& OI & 4.655332 & $17.94_{-2.12}^{+2.35}$ & $14.05_{-0.31}^{+0.14}$ & component-a  \\
&    & 4.655690 & $17.72_{-1.98}^{+2.30}$ & $14.93_{-0.09}^{+0.12}$ & component-b  \\
&    & 4.656426 & $4.65_{-1.06}^{+1.66}$ & $16.11_{-1.00}^{+0.93}$ & component-c  \\
&    & 4.656916 & $11.65_{-2.43}^{+2.18}$ & $14.05_{-0.08}^{+0.07}$ & component-d  \\
& SiII & 4.655332 & $17.94_{-2.12}^{+2.35}$ & $11.17_{-0.79}^{+0.87}$ & component-a  \\
&    & 4.655690 & $17.72_{-1.98}^{+2.30}$ & $14.06_{-0.04}^{+0.04}$ & component-b  \\
&    & 4.656426 & $4.65_{-1.06}^{+1.66}$ & $12.76_{-1.25}^{+0.34}$ & component-c  \\
&    & 4.656916 & $11.65_{-2.43}^{+2.18}$ & $13.04_{-0.53}^{+0.22}$ & component-d  \\
& CII & 4.655332 & $17.94_{-2.12}^{+2.35}$ & $13.94_{-0.09}^{+0.08}$ & component-a  \\
&    & 4.655690 & $17.72_{-1.98}^{+2.30}$ & $14.11_{-0.08}^{+0.08}$ & component-b  \\
&    & 4.656426 & $4.65_{-1.06}^{+1.66}$ & $15.20_{-0.81}^{+1.11}$ & component-c  \\
&    & 4.656916 & $11.65_{-2.43}^{+2.18}$ & $13.82_{-0.08}^{+0.08}$ & component-d  \\

\end{longtable}
\end{center}

\begin{figure}[ht!]
    \centering
    \includegraphics[width=0.8\linewidth]{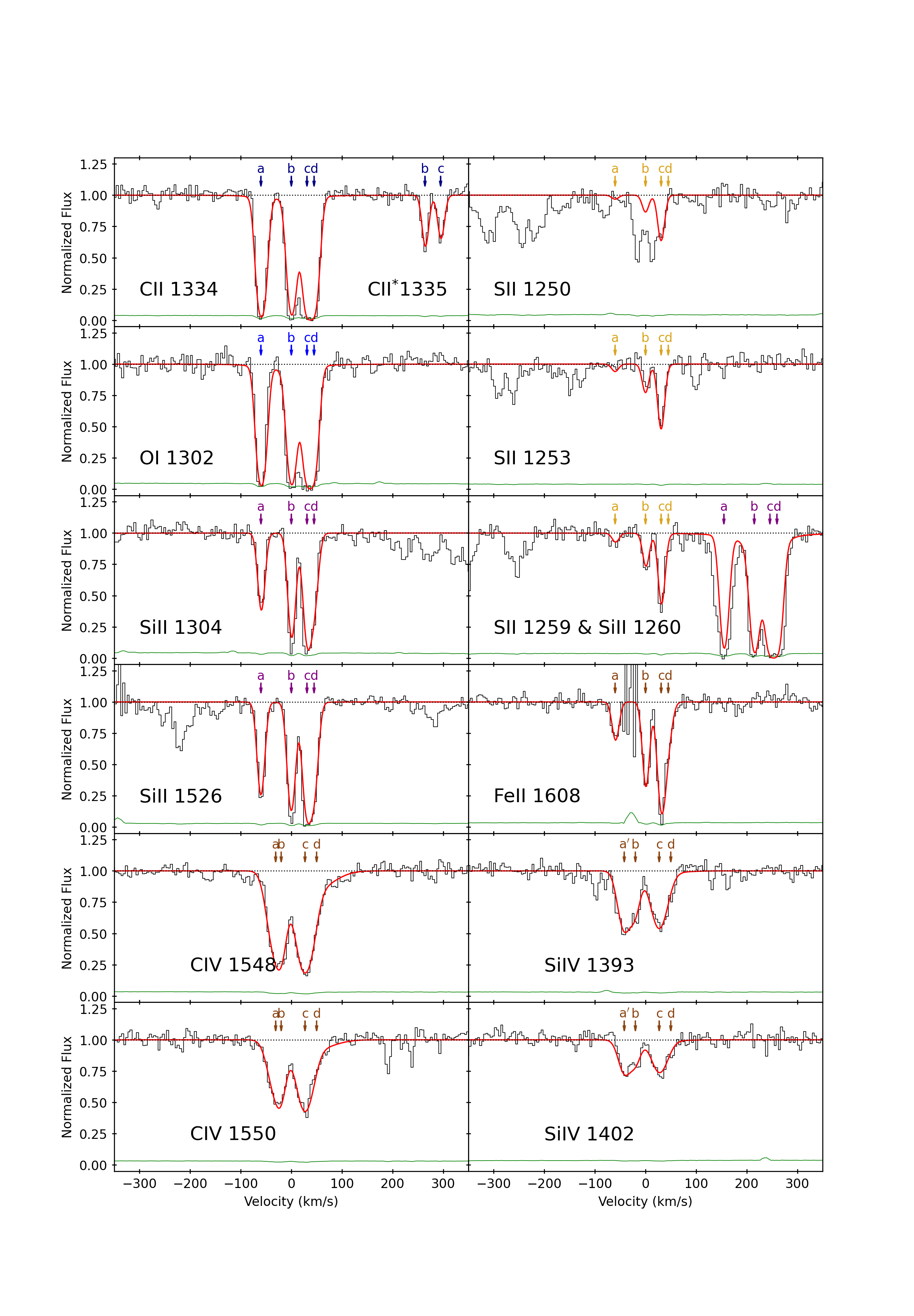}
    \caption{Voigt profiles of major metal lines from our MIKE data for the DLA along J0007 sightline at $z_{DLA}=4.24168$. The normalized quasar spectra are shown in black, the fitted profiles are shown in solid red, and the 1 $\sigma$ uncertainties in the normalized flux are shown in green. The arrows indicate the components of the absorber.
    \label{fig:J0007_metals}}
\end{figure}

\begin{figure}[ht!]
    \centering
    \includegraphics[width=\linewidth]{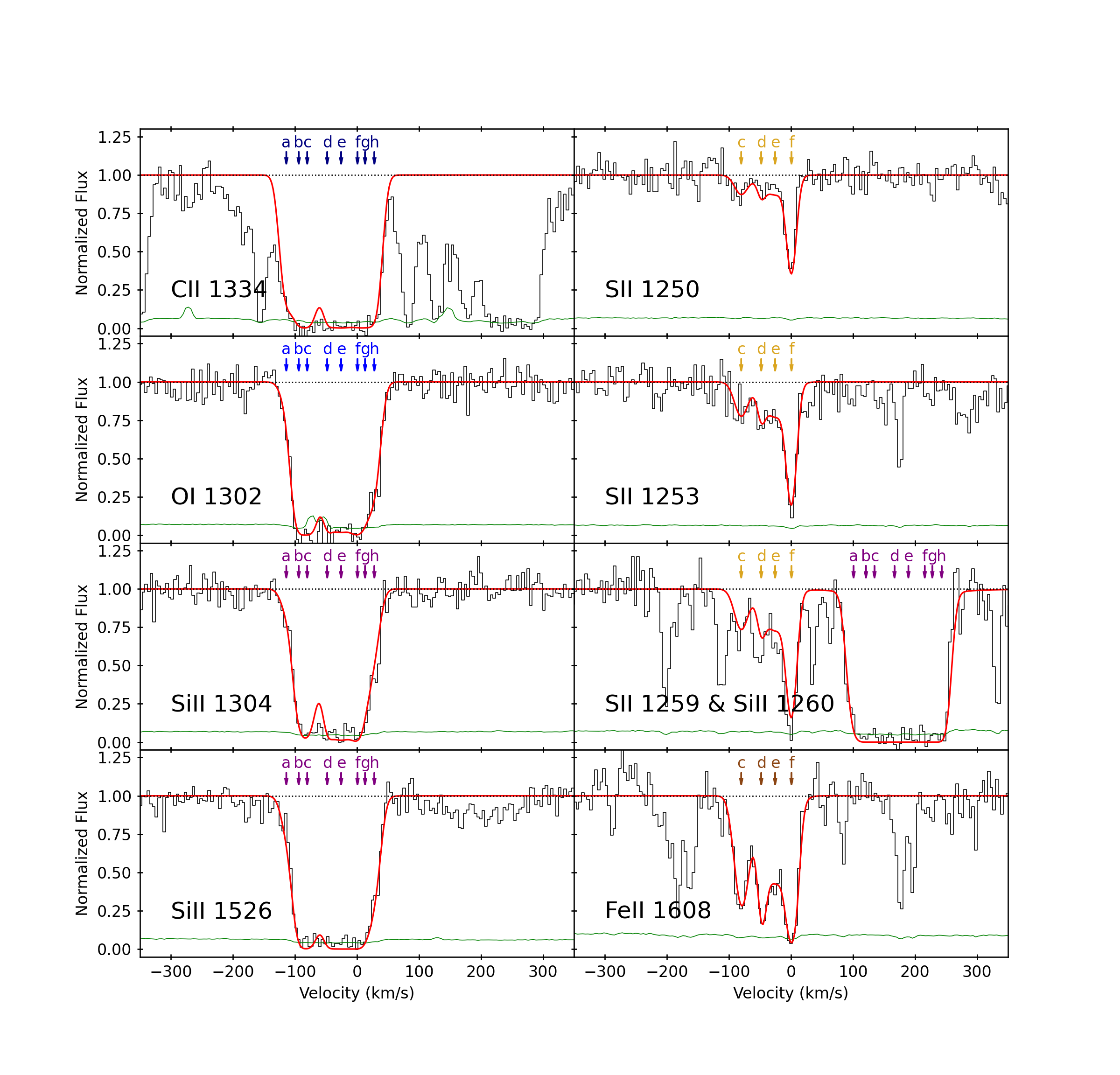}
    \caption{Voigt profiles of major metal lines for the DLA along J0838 sightline at $z_{DLA}=4.713615$.
    \label{fig:J0838_metals}}
\end{figure}

\begin{figure}[ht!]
    \centering
    \includegraphics[width=\linewidth]{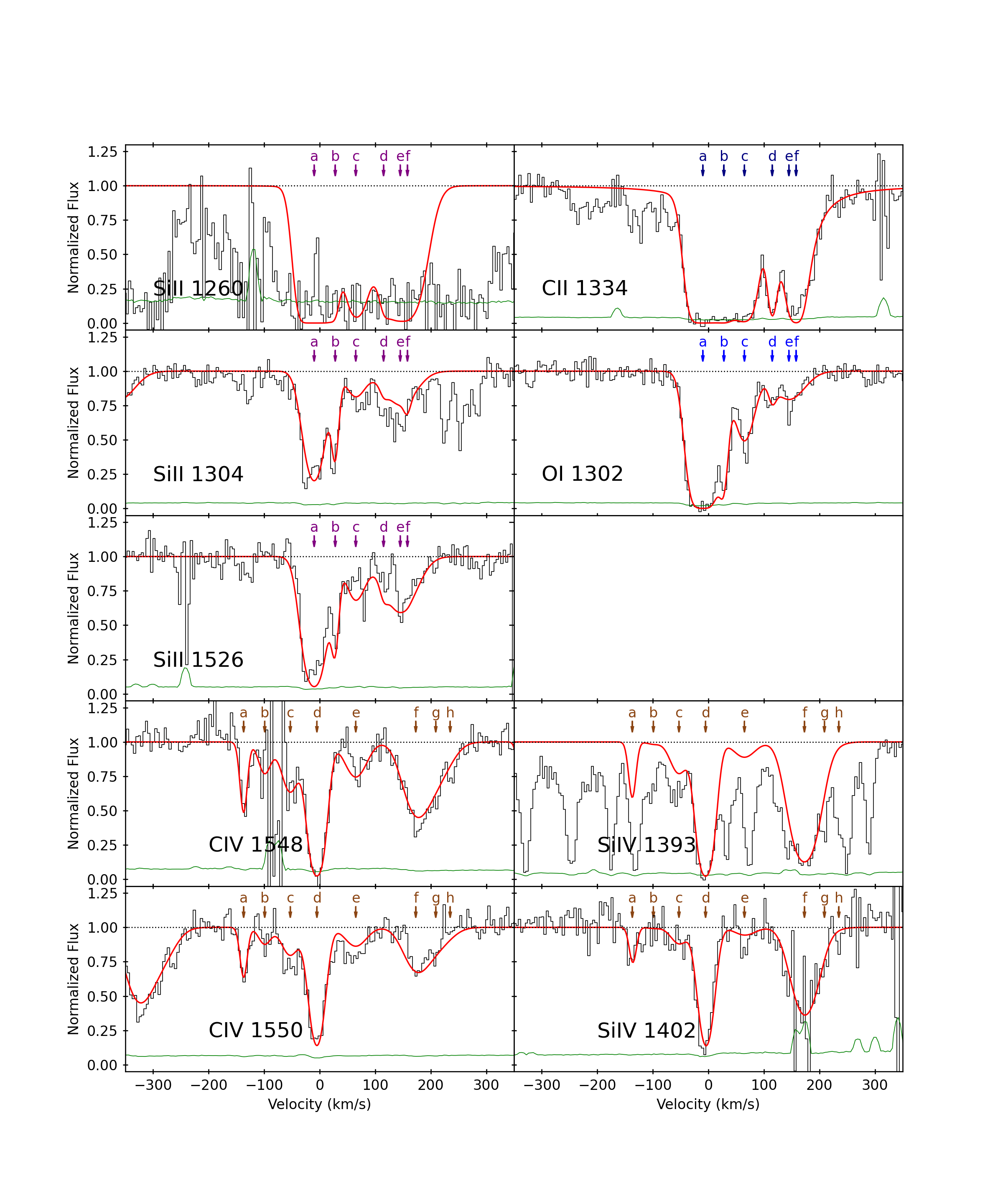}
    \caption{Voigt profiles of major metal lines for the DLA along J1129 sightline at $z_{DLA}=4.494940$.
    \label{fig:J1129_metals}}
\end{figure}

\begin{figure}[ht!]
    \includegraphics[width=\linewidth]{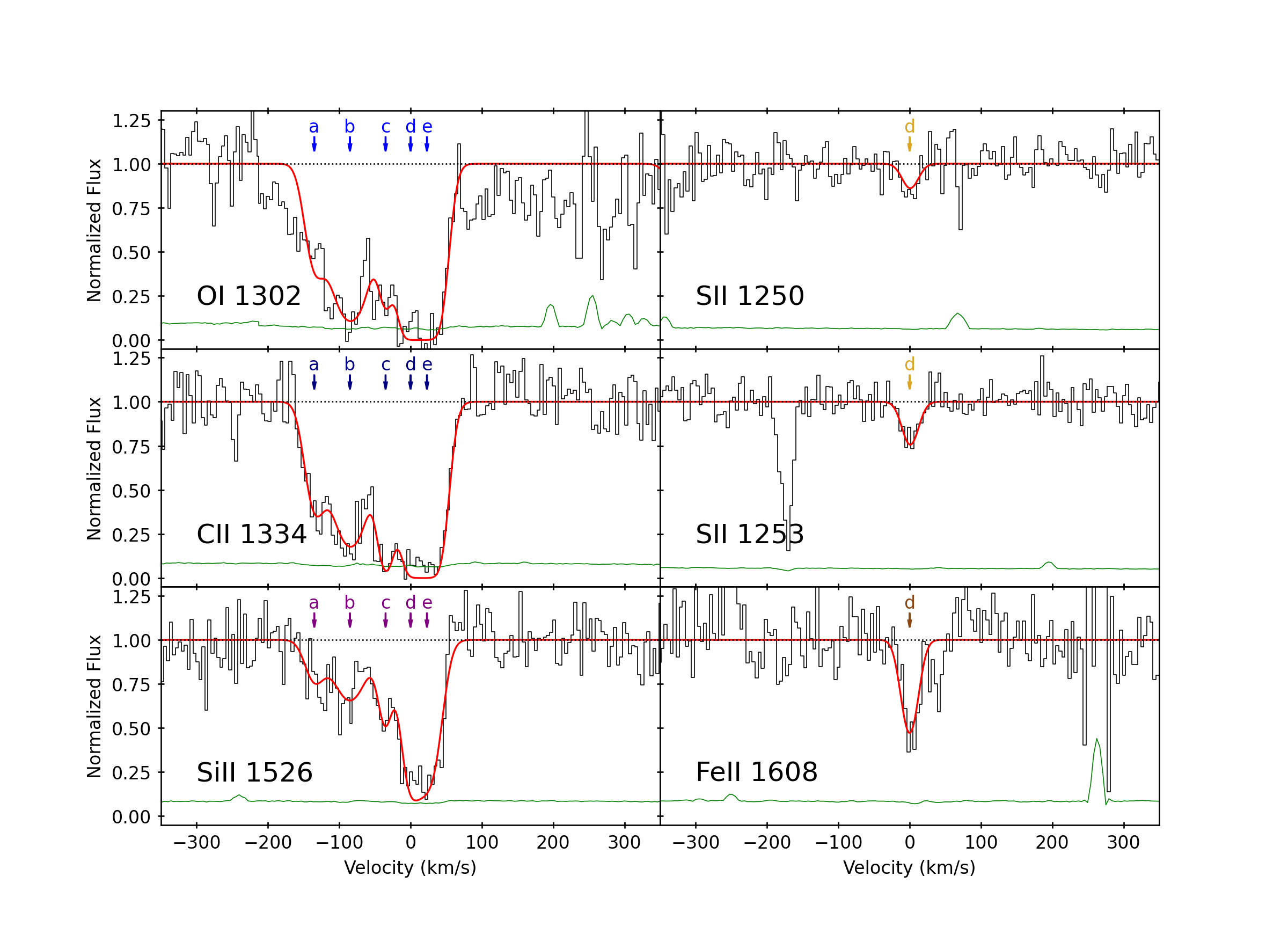}
    \caption{Voigt profiles of major metal lines for the DLA along J1135 sightline at $z_{DLA}=4.555380$.
    \label{fig:J1135_metals}}
\end{figure}

\begin{figure}[ht!]
    \includegraphics[width=\linewidth]{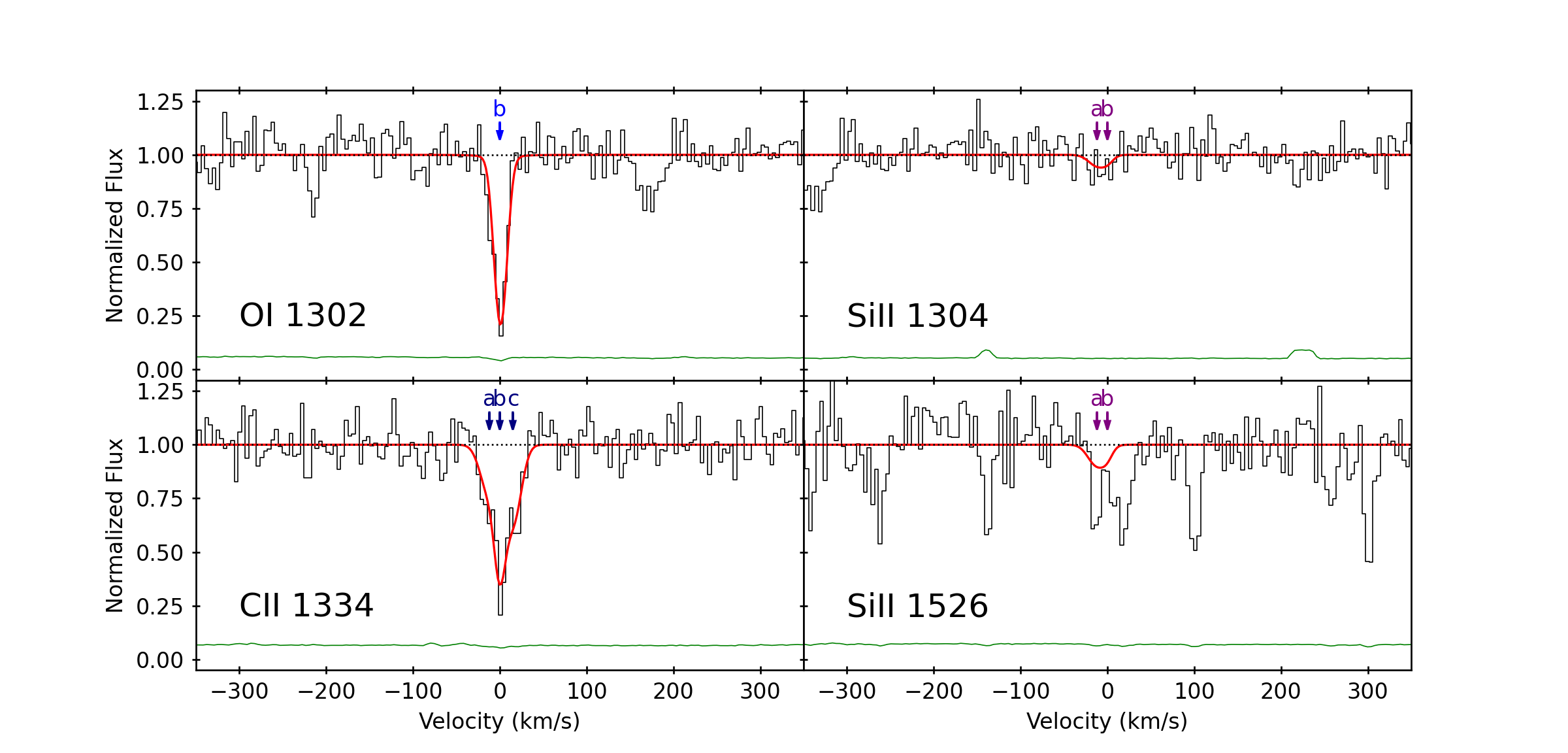}
    \caption{Voigt profiles of major metal lines for the DLA along J1135 sightline at $z_{DLA}=4.346010$.
    \label{fig:J1135_metals-lowz}}
\end{figure}

\begin{figure}[ht!]
    \includegraphics[width=\linewidth]{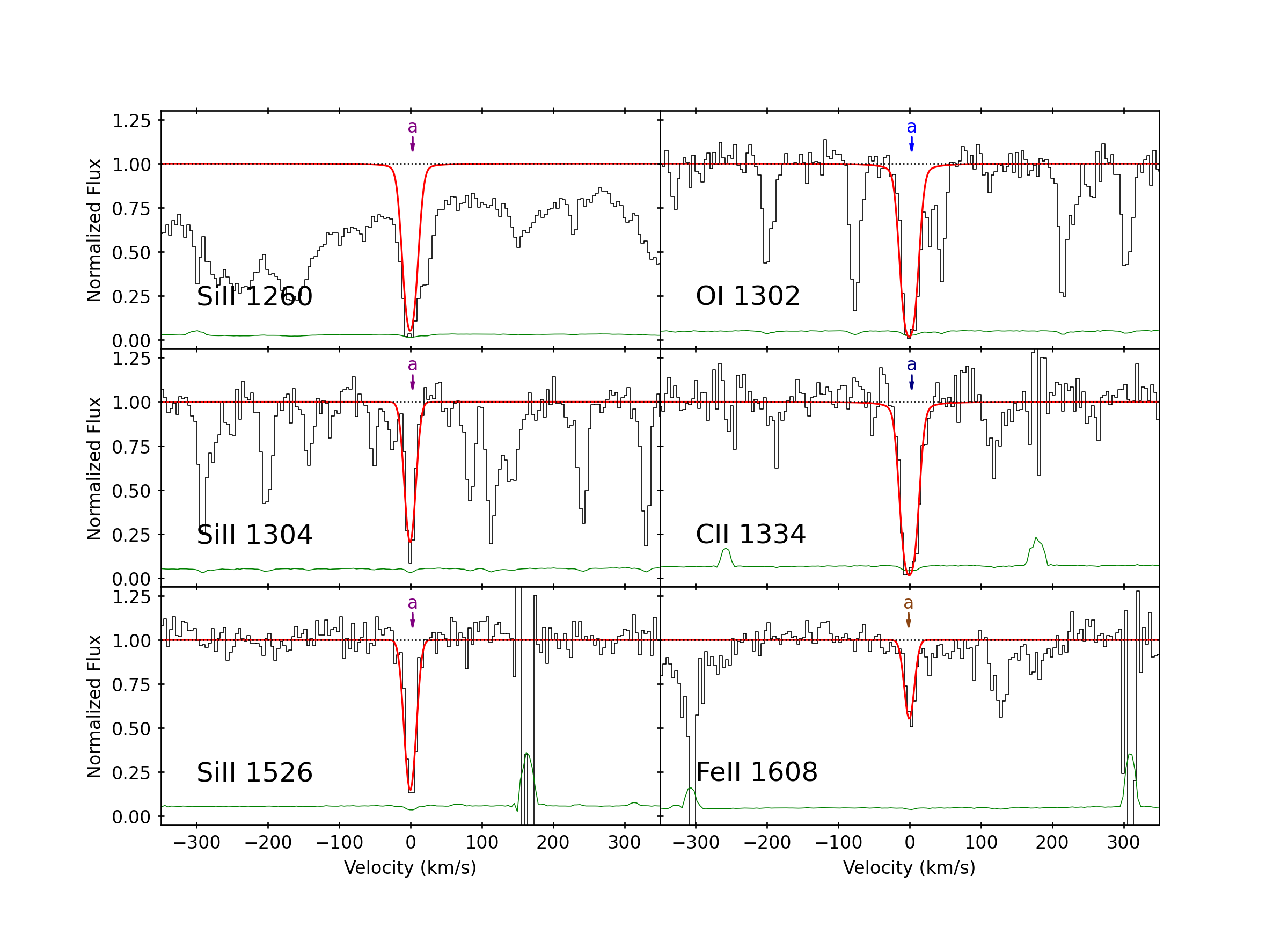}
    \caption{Voigt profiles of major metal lines for the DLA along J1336 sightlineat $z_{DLA}=4.518762$.
    \label{fig:J1336_metals}}
\end{figure}

\begin{figure}[ht!]
    \includegraphics[width=\linewidth]{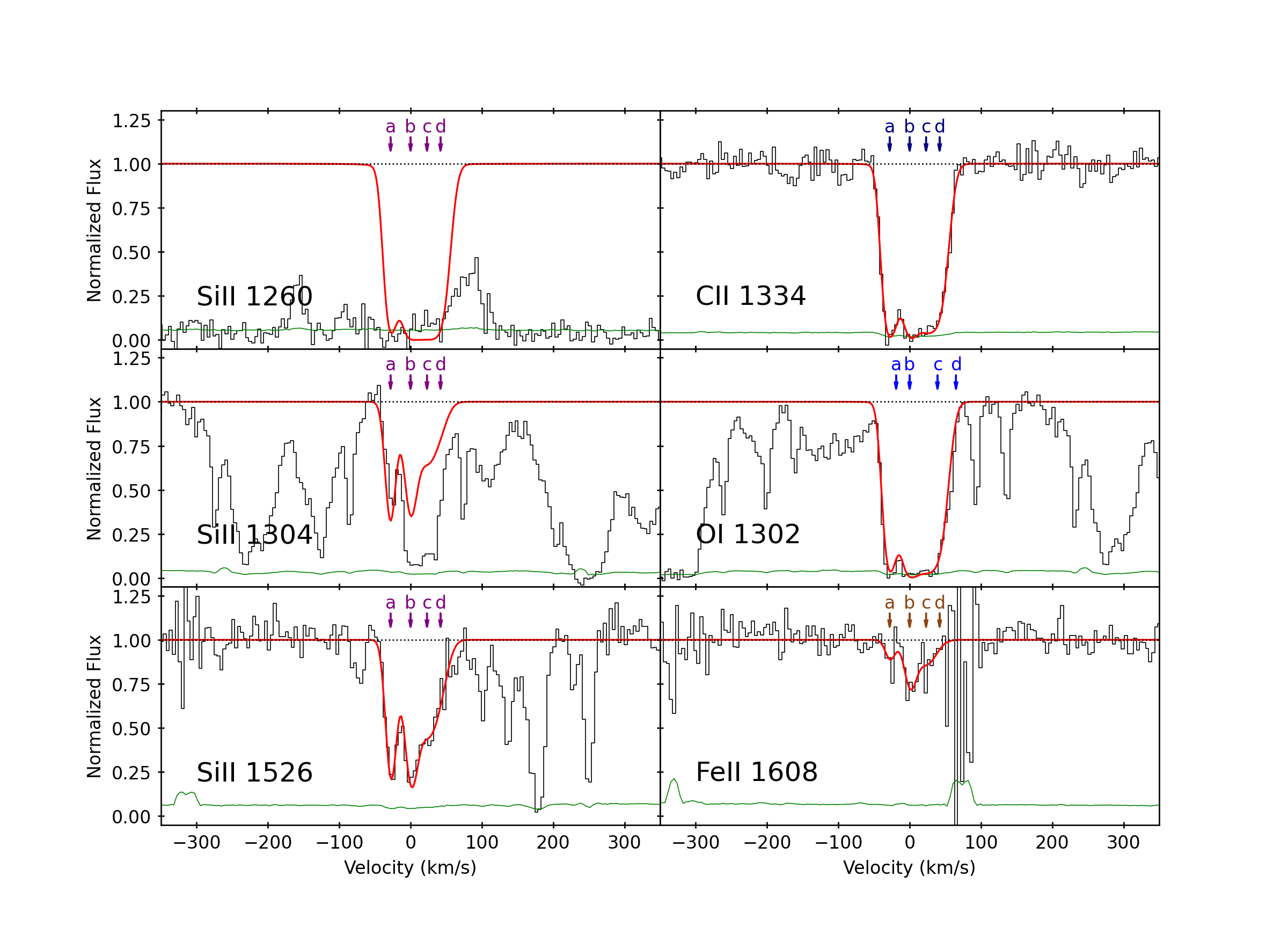}
    \caption{Voigt profiles of major metal lines for the DLA along J1336 sightline at $z_{DLA}=4.286280$.
    \label{fig:J1336_metals-lowz}}
\end{figure}

\begin{figure}[ht!]
    \centering
    \includegraphics[width=\linewidth]{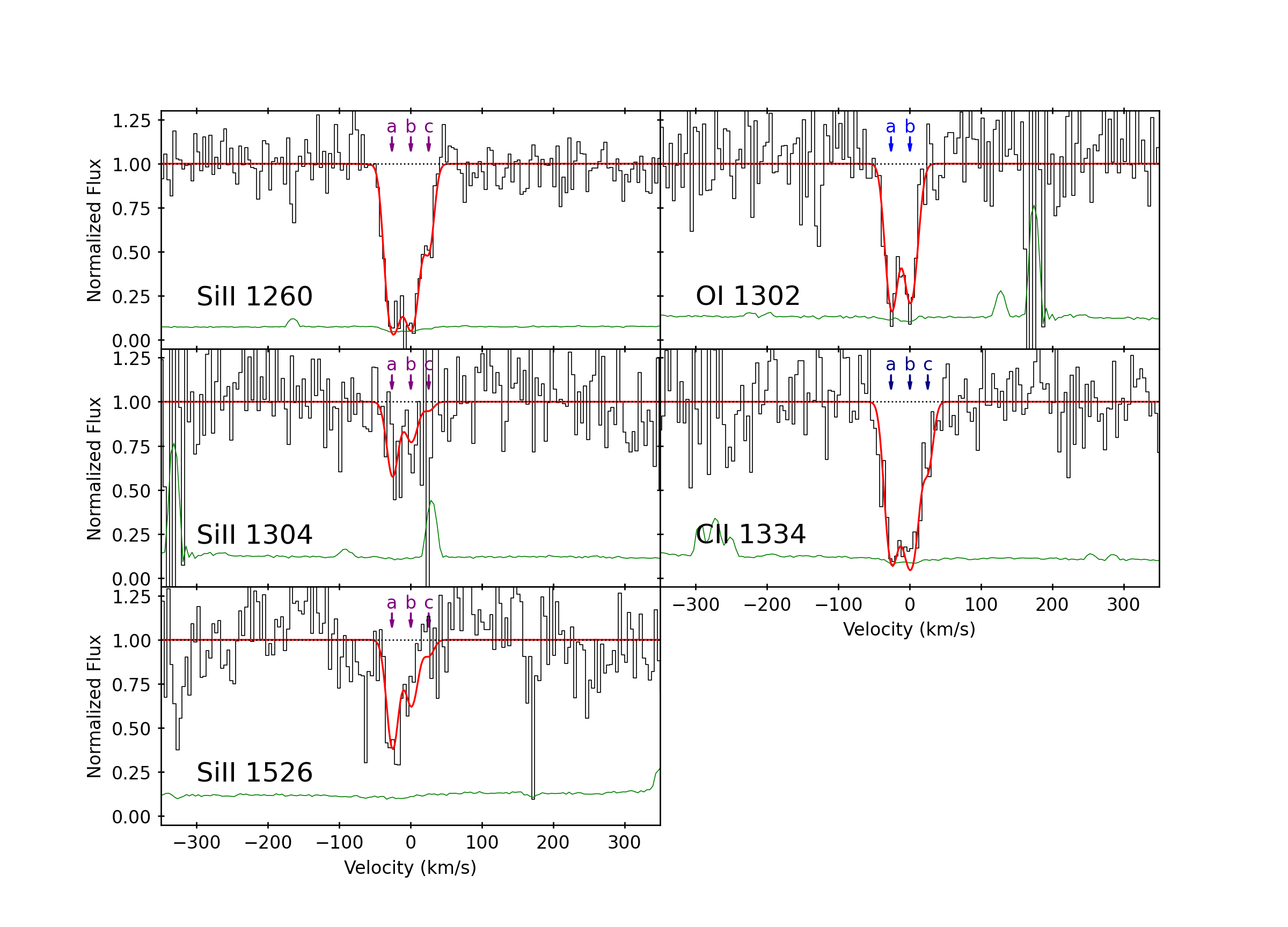}
    \caption{Voigt profiles of major metal lines for the DLA along J2202 sightline at $z_{DLA}=4.948650$.
    \label{fig:J2202_metals}}
\end{figure}

\begin{figure}[ht!]
    \centering
    \includegraphics[width=\linewidth]{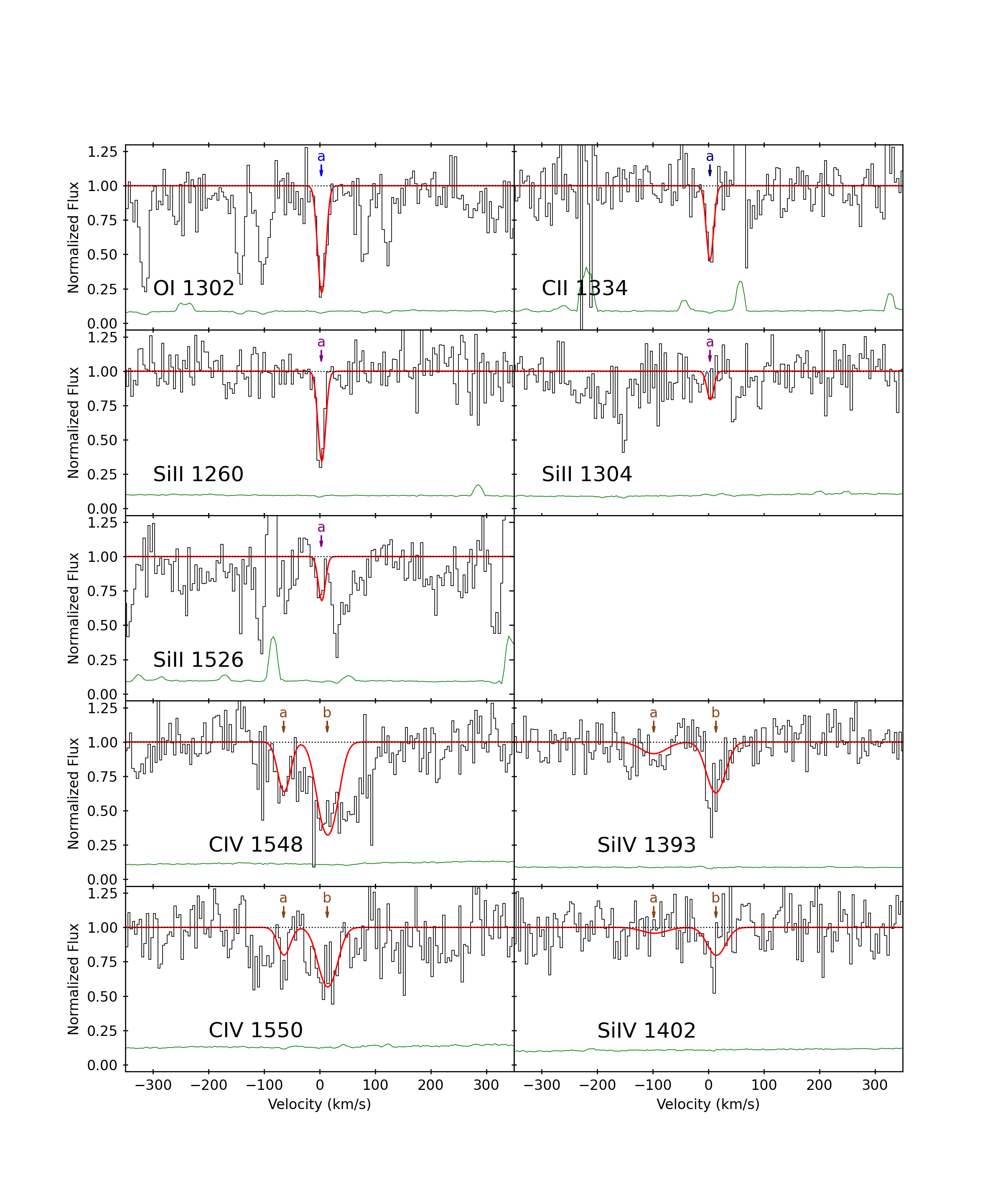}
    \caption{Voigt profiles of major metal lines for the DLA along J2328 sightline at $z_{DLA}=4.889080$.
    \label{fig:J2328_metals}}
\end{figure}

\begin{figure}[ht!]
    \centering
    \includegraphics[width=\linewidth]{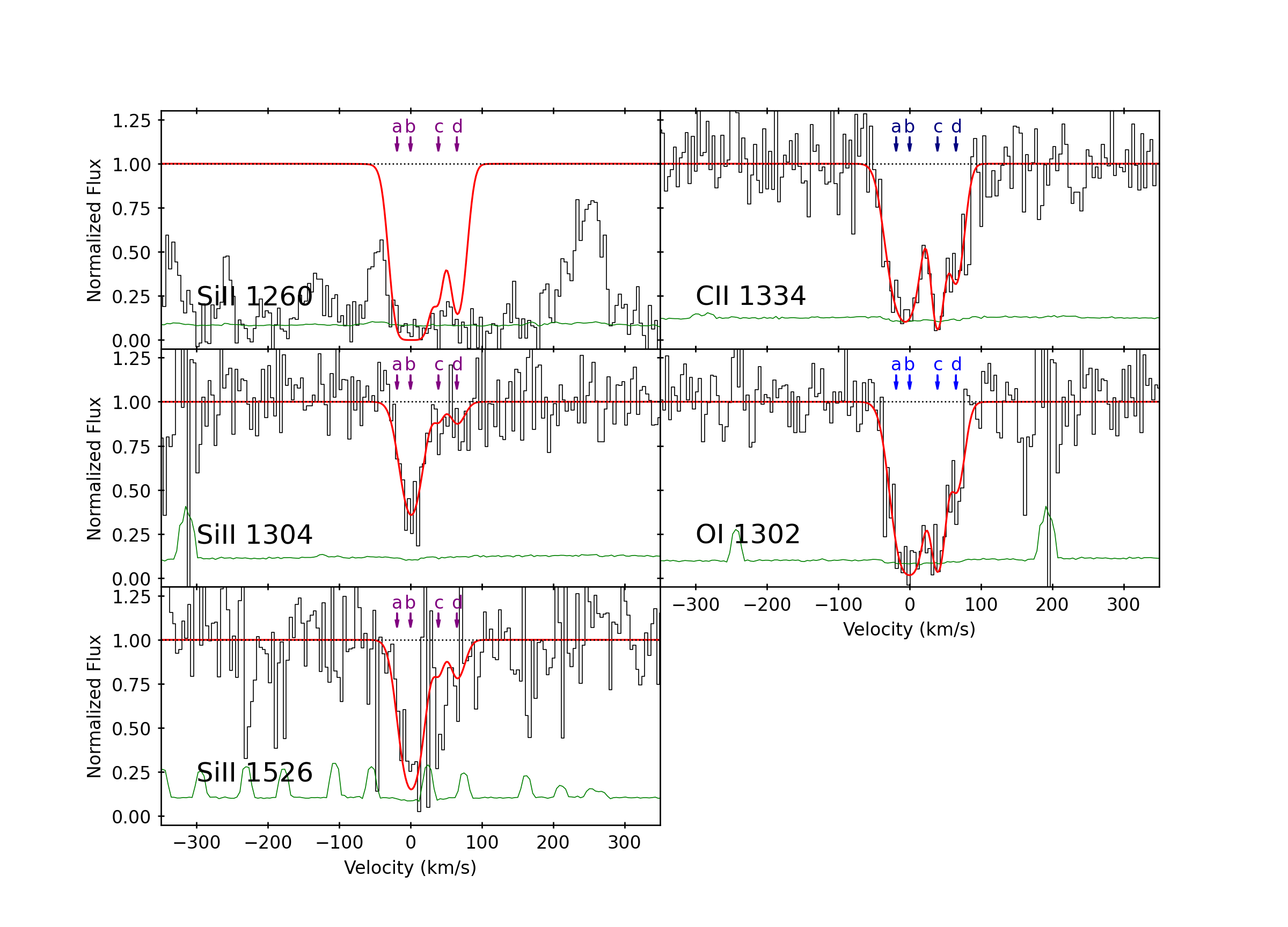}
    \caption{Voigt profiles of major metal lines for the DLA along J2328 sightline at $z_{DLA}=4.655690$.
    \label{fig:J2328_metals-lowz}}
\end{figure}

\begin{figure}
    \centering
    \includegraphics[scale=1.0]{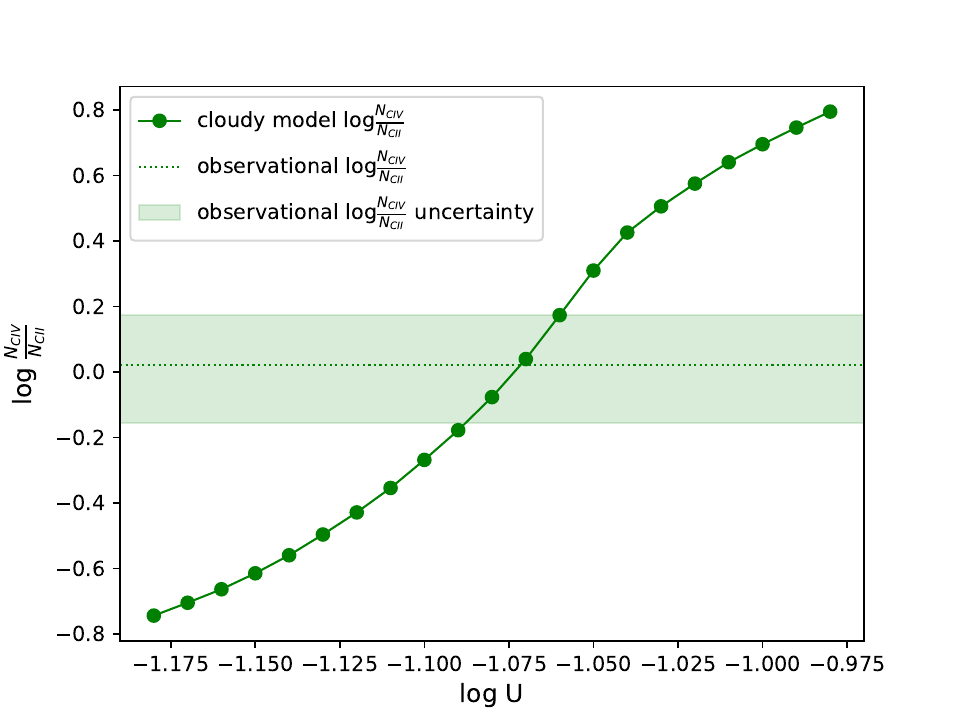}
    \caption{Ionization parameter (log$U$) as a function of raltive ratios of log$\frac{N_{CIV}}{N_{CII}}$ in {\tt cloudy} models for the sub-DLA along J2328+0217 at $z_{DLA} = 4.889080$. From observed log$\frac{N_{CIV}}{N_{CII}}$, ioniazation parameter is constrained $-1.09<$log$U<-1.06$. }
    \label{fig:photonionization}
\end{figure}

\vspace{5mm}
\facilities{Magellan (MIKE)}

\software{astropy \citep{2013A&A...558A..33A,2018AJ....156..123A},  
          Cloudy \citep{2013RMxAA..49..137F} 
          }

\end{document}